\overfullrule=0pt
\input harvmac
\def\bar{\overline}
\def\a{{\alpha}}

\def\ah{{\widehat \a}}
\def\lh{{\widehat \lambda}}

\def\wh{{\widehat w}}

\def\l{{\lambda}}

\def\lt{{\lambda '}}
\def\lbt{{\bar\lambda '}}
\def\At{{A'}}

\def\Ht{{\widetilde H}}
\def\Abt{{\bar A'}}
\def\Pbt{{\bar P'}}
\def\wt{{w'}}
\def\wbt{{\bar w'}}

\def\b{{\beta}}
\def\bh{{\widehat\beta}}

\def\g{{\gamma}}

\def\d{{\delta}}
\def\e{{\epsilon}}
\def\s{{\sigma}}
\def\k{{\kappa}}
\def\kb{{\bar\kappa}}

\def\L{{\Lambda}}
\def\O{{\Omega}}
\def\half{{1\over 2}}
\def\p{{\partial}}
\def\dt{{\partial\over{\partial\tau}}}

\def\t{{\theta}}
\def\md{{\dot\mu}}
\def\nd{{\dot\nu}}
\def\T{{\Theta}}
\def\S{{\Sigma}}
\def\th{{\widehat\theta}}
\def\Tb{{\bar\Theta}}
\def\tb{{\bar\theta}}

\Title{\vbox{\baselineskip12pt
\hbox{IFT-P.06/2007 }}}
{{\vbox{\centerline{ A New Limit of the 
$AdS_5\times S^5$ Sigma Model}}} }
\bigskip\centerline{Nathan Berkovits\foot{e-mail: nberkovi@ift.unesp.br}}
\bigskip
\centerline{\it Instituto de F\'\i sica Te\'orica, State University of
S\~ao Paulo}
\centerline{\it Rua Pamplona 145, 01405-900, S\~ao Paulo, SP, Brasil}

\vskip .3in

Using the pure spinor formalism, a quantizable sigma model has been
constructed for the superstring in an $AdS_5\times S^5$ background
with manifest $PSU(2,2|4)$ invariance. The $PSU(2,2|4)$ metric $g_{AB}$
has both vector components $g_{ab}$ and spinor components 
$g_{\alpha\beta}$, and in the limit where the spinor components 
$g_{\alpha\beta}$
are taken to infinity, the $AdS_5\times S^5$ sigma model reduces to
the worldsheet action in a flat background.

In this paper, we instead consider the limit where the vector components
$g_{ab}$ are taken to infinity. In this limit, the 
$AdS_5\times S^5$ sigma model simplifies to a topological A-model 
constructed from fermionic N=2 superfields whose bosonic components
transform like twistor variables. Just as d=3 Chern-Simons
theory can be described by the open string sector of a topological A-model,
the open string sector of this topological A-model
describes d=4 N=4 super-Yang-Mills. These results might be useful
for constructing a worldsheet proof of the Maldacena conjecture
analogous to the Gopakumar-Vafa-Ooguri worldsheet proof of 
Chern-Simons/conifold duality.

\vskip .3in

\Date {March 2007}

\newsec{Introduction}

Maldacena's conjecture \ref\malda{J. Maldacena,
{\it The large N limit of superconformal field theories and
supergravity}, Adv. Theor. Math. Phys. 2 (1998) 231, hep-th/9711200.}
relating d=4 N=4 super-Yang-Mills and
the superstring on $AdS_5\times S^5$ has been verified in various
limiting cases. However, in the limit where d=4 N=4 super-Yang-Mills
is weakly coupled, it has been difficult to verify the conjecture because
the $AdS_5\times S^5$ background is highly curved. 
Although there exists a quantizable sigma model description of the
superstring in an $AdS_5\times S^5$ background using the pure spinor
formalism \ref\purer{N. Berkovits,
{\it Super-Poincar\'e covariant quantization of the superstring},
JHEP 0004 (2000) 018, hep-th/0001035.}, 
the sigma model naively becomes strongly coupled when the 
$AdS_5\times S^5$ radius goes to zero.

In an $AdS_5\times S^5$ background, the sigma model action using
the pure spinor formalism has the form \purer\ref\adsv{N. Berkovits and
O. Chandia, {\it Superstring vertex operators in an $AdS_5\times S^5$
background}, Nucl. Phys. B596 (2001) 185, hep-th/0009168.}
\ref\vall{
B.C. Vallilo, {\it One-loop conformal invariance of the superstring
in an $AdS_5\times S^5$ background}, JHEP 0212 (2002) 042, hep-th/0210064.}
\ref\adscons{N. Berkovits, {\it Quantum consistency of the superstring
in $AdS_5\times S^5$ background}, JHEP 0503 (2005) 041, hep-th/0411170.}
\eqn\sone{S = {1\over \Lambda}
\int d^2 z
[\half \eta_{ab} J^a \bar J^b + \eta_{\a\bh}( {3\over 4} J^\bh\bar J^\a
- {1\over 4} \bar J^\bh J^\a) + {\rm  ghost~~contribution} ] }
where 
$J^a$ for $a=0$ to 9 and $(J^\a, J^\bh)$ for $\a,\bh=1$ to 16
are bosonic and fermionic ${PSU(2,2|4)}\over{SO(4,1)\times SO(5)}$
currents constructed from the worldsheet Green-Schwarz
variables $(x, \t,\th)$ 
as in the Metsaev-Tseytlin construction 
\ref\metsaev{R.R. Metsaev and A.A. Tseytlin
{\it Type IIB superstring action in $AdS_5\times S^5$ background},
Nucl. Phys. B533 (1998) 109, hep-th/9805028.}, $\eta_{ab}$
is the d=10 Minkowski metric and $\eta_{\a\bh}
= (\g^{01234})_{\a\bh}$. BRST invariance together with $PSU(2,2|4)$ 
invariance uniquely fixes the relative coefficients in the action, so
the $AdS_5\times S^5$ radius $r$ 
only appears in the action through the sigma model
coupling constant $\Lambda= \alpha'/r^2$ 
where $\a'$ is the inverse string tension. So the sigma model
seems to be strongly coupled when  
the $AdS_5\times S^5$ radius is small. However, this conclusion may be too naive
since it assumes that the $PSU(2,2|4)$ algebra remains undeformed
when the $AdS_5\times S^5$ radius is taken to zero.

One limit of the sigma model which is well-understood is the d=10 flat
space limit where the $AdS_5\times S^5$ radius goes to infinity. Naively, one
would go to the flat space limit by simply taking $\Lambda\to 0$, however,
this limit would preserve $PSU(2,2|4)$ invariance instead of the desired
d=10 super-Poincar\'e invariance. The correct way to go to the flat
space limit is to rescale
the spinor component of the $PSU(2,2|4)$
metric $g_{\a\bh} = \eta_{\a\bh}$ to
\eqn\rescaleone{g_{\a\bh} = r \eta_{\a\bh}}
in the sigma model action of \sone, together with
an appropriate rescaling of the $PSU(2,2|4)$ structure constaints. In the limit
where $r$ goes to infinity, the $PSU(2,2|4)$ algebra is deformed into
the d=10 super-Poincar\'e algebra and the second-order kinetic term for
the fermions in \sone\ blows up.
Nevertheless, this limit can be 
taken smoothly by writing the second-order kinetic
term 
$r \eta_{\a\bh} J^\bh \bar J^\a$ as the first-order kinetic term
$\bar J^\a d_\a + J^\bh \widehat d_\bh + r^{-1}\eta^{\a\bh}
d_\a \widehat d_\bh$ where $d_\a$ and $\widehat d_\bh$ are auxiliary fermionic
variables. In the limit where $r\to \infty$, one obtains a first-order action
for the worldsheet fermions $(\t^\a,d_\a)$ and $(\th^\bh, \widehat d_\bh)$,
which is the flat space version of the worldsheet action using the pure
spinor formalism.

Since the structure constants of the algebra 
are related
to the superspace torsions $T_{AB}{}^{C}$, this limiting procedure
can be understood as a rescaling of the $AdS_5\times S^5$ superspace
torsions into the flat superpace torsions.
In an $AdS_5\times S^5$ background, 
$T_{\a a}{}^\bh$ and $T_{\a \b}{}^a$ are non-vanishing torsions which
are related by $T_{\a a}{}^\bh \eta_{\b\bh} = T_{\a\b}{}^b \eta_{ab}$.
On the other hand, in a flat background, 
$T_{\a\b}{}^a$ is non-vanishing and
$T_{\a a}{}^\bh=0$. The rescaling of the structure constants
and $g_{\a\bh}$ as in \rescaleone\
rescales the torsions such
that 
\eqn\ratio {{{T_{\a \b}{}^b \eta_{ab}}\over{T_{\a a}{}^\bh \eta_{\b\bh}}} = r.}
So when $r\to \infty$, $T_{\a a}{}^\bh\to 0$ which corresponds to
flat space.

In this paper, we will consider a different limit of the $AdS_5\times S^5$
sigma model in which, instead of the spinor component of the
$PSU(2,2|4)$ metric $g_{\a\bh}$ 
being rescaled, the vector component $g_{ab}$ will be rescaled as
\eqn\rescaletwo{g_{ab}= r^{-1} \eta_{ab}.} 
Furthermore, the $PSU(2,2|4)$ structure
constants will be rescaled such that in the limit where $r\to 0$,
the $PSU(2,2|4)$ superalgebra
is deformed into an $SU(2,2)\times SU(4)$ bosonic algebra with
32 abelian fermionic symmetries.
This corresponds to rescaling the torsions such that \ratio\ remains
satisfied when $r\to 0$, which implies that 
the resulting background has non-vanishing
$T_{\a a}{}^\bh$ but has $T_{\a\b}{}^a=0$. Since the usual 
construction of supergravity backgrounds assumes that $T_{\a\b}{}^a =
\gamma_{\a\b}^a$ 
\ref\howew{P. Howe and P. West, {\it The complete N=2 D=10 supergravity},
Nucl. Phys. B238 (1984) 181.}, 
this $r\to 0$ limit does not correspond to a standard
supergravity background.

Nevertheless, the resulting sigma model action when $T_{\a\b}{}^a\to 0$ is very
simple and can be expressed as a linear N=2 sigma model constructed from
16 chiral and antichiral N=2 superfields denoted by 
$\T^{r j}$ and $\Tb_{j r}$, where $r=1$ to 4 are $SU(2,2)$ indices
and $j=1$ to 4 are $SU(4)$ indices. Unlike the bosonic superfields
in standard N=2 sigma models, 
$\T^{r j}$ and $\Tb_{j r}$ are fermionic superfields. It is interesting
that in the open-closed matrix model duality of \ref\rastel
{D. Gaiotto and L. Rastelli, {\it A paradigm of open/closed
duality: Liouville D-branes and the Kontsevich model}, JHEP 0507 (2005)
053, hep-th/0312196.}, the matter variables are also described by
fermions with a second-order kinetic action. The lowest
components of $\T^{r j}$ and $\Tb_{jr}$
are linear combinations of the $\t$ and $\th$
variables, and the bosonic components of $\T^{r j}$ and $\Tb_{j r}$
are twistor-like combinations of
the ten $x$'s and
22 pure spinor ghosts. 
Just as the fermionic variables had a first-order kinetic action
in the flat space
sigma model obtained by rescaling \rescaleone, the bosonic variables
now have a first-order kinetic action in the N=2 sigma model obtained
by rescaling \rescaletwo.

Moreover, this N=2 sigma model is twisted as an A-model where the pure 
spinor BRST operator from the original $AdS_5\times S^5$ sigma model
acts in the usual topological manner as the scalar worldsheet supersymmetry
generator. So the N=2 sigma model is a topological A-model with the
worldsheet action 
\eqn\stwo{S = \int d^2 z d^4\k  ~\Tb_{jr} \T^{r j}}
where $(\k_+,\kb_+,\k_-, \kb_-)$ are the Grassmann parameters of
the N=(2,2) superspace.
This model is invariant under the bosonic isometries $SU(2,2)\times SU(4)\times
U(1)$ which act on the superfields as
\eqn\acta{\d\T^{r j} = i \L^r_s \T^{s j} + i \T^{r k} \O_k^j + i \S \T^{r j},
\quad
\d\Tb_{j r} =- i \Tb^{j s}\L^s_r - i \O_j^k\Tb_{k r} - i \S \Tb_{j r},}
where $(\L^r_s, \Omega_j^k, \S)$
are constant parameters satisfying
$\L^r_r = \O^j_j=0$, and is invariant
under the 32 abelian fermionic isometries
\eqn\actb{\d\T^{r j} = \a^{r j}, \quad
\d\Tb_{j r} = \bar\a_{j r} }
where $\a^{r j}$ and $\bar\a_{j r}$ are constant Grassmann parameters.
Note that the bosonic isometries of this model include a ``bonus''
$U(1)$ symmetry \ref\intril{K. Intriligator, {\it
Bonus symmetries of N=4 super-Yang-Mills correlation functions
via AdS duality}, Nucl. Phys. B551 (1999) 575, hep-th/9811047.}
in addition to the $SU(2,2)\times SU(4)$ isometries
of the original $AdS_5\times S^5$ sigma model.

Introducing fermionic worldsheet superfields whose bosonic components
are twistor-like coordinates has been useful in classical descriptions
of the superstring where kappa-symmetry is replaced by worldsheet
supersymmetry \ref\sorokin{D. Sorokin, V. Tkach, D. Volkov
and A. Zheltukhin, {\it From the superparticle Siegel symmetry
to the spinning particle proper time supersymmetry}, Phys. Lett.
B216 (1989) 302.}\ref\matone{
M. Matone, L. Mazzucato,
I. Oda, D. Sorokin and M. Tonin, {\it The superembedding origin of the
Berkovits pure spinor covariant quantization of superstrings},
Nucl. Phys. B639 (2002) 182, hep-th/0206104.}\ref\prsor{
D. Sorokin, {\it Superbranes and superembeddings}, Phys. Rept. 329
(2000) 1, hep-th/9906142.}.
The N=2 model in this paper shares many features with this
``super-embedding'' approach, however, it has the advantage of
being quantizable because of the second-order action for the fermionic
superfields. Since the second-order action for fermionic superfields is
generated by the Ramond-Ramond background, it might be
possible to generalize the twistor-like methods of this paper to
more general Ramond-Ramond backgrounds.

The abelianization of the fermionic isometries of \actb\ comes from
setting $T_{\a\b}{}^a=0$ and means that the supersymmetry
generators anticommute with each other. To relate
this model to super-Yang-Mills where 
supersymmetry acts in the conventional
way, it is useful to interpret \stwo\ as the limit
of a non-linear topological A-model which is constructed
such that the isometries of \acta\ and \actb\ are deformed into
$SU(2,2|4)$ isometries.

The worldsheet action for this non-linear topological A-model is
\eqn\acttwo{S = {1\over \L} \int d^2 z d^4 \k
[ \Tb_{r j}\T^{j r} - {1\over {2R^2}} \Tb_{r j}\T^{j s}\Tb_{s k}\T^{kr}
+ {1\over {3R^4}}
\Tb_{r j}\T^{j s}\Tb_{s k}\T^{k t} \Tb_{t l}\T^{l r} + ... ] }
$$ = {R^2\over \L} \int d^2 z d^4 \k  ~ Tr [ \log (1 + {1\over {R^2}}
\Tb \T) ]$$
where $R$ is a new parameter which, in the limit $R\to \infty$,
takes the non-linear sigma model into the linear sigma model of
\stwo. This non-linear action will be shown to be
one-loop conformally invariant, and is invariant under the same
$SU(2,2)\times SU(4)\times U(1)$ transformations as \acta. But
the fermionic transformations of \actb\ are modified to 
\eqn\actc{\d\T^{r j} = \a^{r j} + {1\over R^2}
\T^{r k}\bar\a_{k s}\T^{s j}, \quad
\d\Tb_{j r} = \bar\a_{j r}  +{1\over R^2} \Tb_{j s}\a^{s k} \Tb_{k r},}
which anticommute to form the superalgebra $SU(2,2|4)$. 

It will be conjectured that the BRST cohomology in the closed string
sector of this non-linear topological A-model is trivial, which implies
that the open string physical states are independent of $R$ and $\L$ in
\acttwo. This
would be similar to the topological A-model for d=3 Chern-Simons which
has physical states only in the open string sector \ref\csw{
E. Witten, {\it Chern-Simons gauge theory as a string theory},
Prog. Math. 133 (1995) 637, hep-th/9207094.},
but would be different from the topological B-model for the 
twistor-string \ref\wittwis{E. Witten,
{\it Perturbative gauge theory as a string theory in twistor
space}, Comm. Math. Phys. 252 (2004) 189, hep-th/0312171.}
which describes N=4 d=4 super-Yang-Mills in the
open sector and N=4 d=4 conformal supergravity in the closed sector.

In the topological A-model for d=3 Chern-Simons, the open
string boundary conditions are $X^I = \bar X_I$ where $X^I$
and $\bar X_I$ are chiral and anti-chiral superfields for $I=1$
to 3. Similarly, the open string boundary conditions in the
non-linear topological A-model of \acttwo\ are $\T^{r j}= 
\Tb_{j r}$. These boundary conditions eliminate half of the
32 $\t$'s and break $SU(2,2|4)$ invariance down to an
$OSp(4|4)$ subgroup, which is the N=4 supersymmetry algebra on
$AdS_4$. In this open topological
A-model, the BRST cohomology of physical states will be shown
to describe d=4 N=4 super-Yang-Mills,
where the bosonic components of $\T^{r j}$ are interpreted
as twistor coordinates constructed from the four $x$'s of
$AdS_4$ together with an N=4 d=4 pure spinor.

The similarities between Chern-Simons and N=4 d=4 super-Yang-Mills
are not surprising since, using the pure spinor formalism, 
the d=10 super-Yang-Mills action can be written in the Chern-Simons
form
$S= \langle V Q V + {2\over 3} V^3\rangle$
where $Q$ is the pure spinor BRST operator and $V$ is the super-Yang-Mills
vertex operator
\ref\spar{N. Berkovits, {\it Covariant quantization
of the superparticle using pure spinors}, JHEP 0109 (2001) 016, 
hep-th/0105050.}\ref\wchw{J.H. Schwarz and E. Witten, private communication.}.
Furthermore, there is a gauge/geometry correspondence relating
Chern-Simons and the resolved conifold which has many features
in common with the Maldacena conjecture relating N=4 d=4
super-Yang-Mills and $AdS_5\times S^5$. The 
Chern-Simons/conifold correspondence was first proposed by
Gopakumar and Vafa \ref\gopv{R. Gopakumar and C. Vafa,
{\it On the gauge theory/geometry correspondence}, Adv. Theor. Math. Phys.
3 (1999) 1415, hep-th/9811131.}, and was later proven using open-closed
duality arguments by Ooguri and Vafa \ref\oov{H. Ooguri and C. Vafa,
{\it Worldsheet derivation of a large N duality}, Nucl. Phys. B641 (2002)
3, hep-th/0205297.}. 

The basic idea behind the open-closed duality proof of
Gopakumar-Vafa-Ooguri is that, in a certain
limit, the closed topological string
theory for the resolved conifold geometry develops
a new branch corresponding to ``holes'' on the closed worldsheet.
These holes were then shown to correspond to the open string
sector of the topological A-model that describes d=3 Chern-Simons.

Since the open string sector of the topological A-model in
this paper describes d=4 N=4 super-Yang-Mills, and since this
topological A-model is related to a certain limit of the closed
superstring in an $AdS_5\times S^5$ background, it is natural
to try to construct a similar open-closed duality proof for the
Maldacena conjecture. However, there are some questions that need
to be answered before such a proof can be attempted.

One question is to explain the interpretation of the torsion ratio
of \ratio\ as the $AdS\times S^5$ radius. Although this interpretation is
easily understood in the flat space limit where $r\to \infty$, 
it is not obvious this interpretation is correct in the limit where
$r\to 0$. So it is not clear that the limit discussed in this paper
corresponds to weak coupling on the super-Yang-Mills side of the 
duality.

A second question is to compute the complete cohomology
of physical states for the topological A-model of \acttwo.
Although it will be shown that the cohomology in the open string
sector of this A-model describes d=4 N=4 super-Yang-Mills, it
remains to be shown that there are no physical states in 
the closed string sector of this A-model.

Finally, a third question which needs to be answered is if
the open string topological A-model in this paper
can be interpreted as a branch
of the closed string $AdS_5 \times S^5$ sigma model which emerges
in the limit where $T_{\a\b}{}^a \to 0$. Perhaps the ``bonus''
$U(1)$ symmetry in \acta\ will play a role in the emergence of this
branch.

In section 2 of this paper, the $AdS_5 \times S^5$
sigma model using the pure spinor formalism is reviewed and
the flat space limit
is discussed. In section 3, the $AdS_5\times S^5$ sigma model
is shown to reduce to a linear topological A-model in the limit
where $T_{\a \b}{}^a\to 0$. In section 4, this linear topological A-model
is deformed into a non-linear topological A-model with $PSU(2,2|4)$
invariance. And in section 5, the open string
sector of this non-linear topological A-model is shown to describe
d=4 N=4 super-Yang-Mills.

\newsec{Review of Pure Spinor Formalism in $AdS_5\times S^5$ Background}

Using the pure spinor formalism, the superstring can be quantized
in any consistent d=10 supergravity background \ref\howe
{N. Berkovits and P. Howe, {\it Ten-dimensional supergravity 
constraints from the pure spinor formalism for the superstring},
Nucl. Phys. B635 (2002) 75, hep-th/0112160.}. Unlike the Green-Schwarz
formalism where the gauge-fixing procedure of kappa-symmetry is
poorly understood even in a flat background, the pure spinor formalism
is quantized using a BRST operator which can be defined in any
consistent supergravity background. 
In an $AdS_5\times S^5 $ background, the BRST transformations act
in a geometric manner, which has been useful for proving the quantum
consistency of this background \adscons.

\subsec{Sigma model action}

The sigma model for the superstring in an $AdS_5\times S^5$
background is manifestly $PSU(2,2|4)$-invariant and is
constructed from the Metsaev-Tseytlin left-invariant currents
\metsaev\
\eqn\MT{J^A = (G^{-1}\p G)^A, \quad
\bar J^A = (G^{-1}\bar\p G)^A, }
where
$G(x,\t,\widehat\t)$ 
takes values in the coset ${PSU(2,2|4)}\over{SO(4,1)\times SO(5)}$,
$A= ([ab],c,\a,\ah)$ ranges over the
30 bosonic and 32 fermionic elements in the Lie algebra of $PSU(2,2|4)$,
$[ab]$ labels the $SO(4,1)\times SO(5)$ ``Lorentz''
generators, $c=0$ to 9 labels
the ``translation'' generators, and $\a,\ah=1$ to 16 label the 
fermionic ``supersymmetry'' generators.
The action in the pure spinor formalism also involves left and
right-moving bosonic ghosts, $(\l^\a, w_\a)$ and $(\lh^\ah,\wh_\ah)$,
which satisfy the pure spinor constraints
$\l\g^c\l=\lh\g^c\lh=0$. Because of the pure spinor constraints,
$w_\a$ and $\widehat w_\ah$ can only appear in combinations which are
invariant under the gauge transformations
\eqn\gaugew{\d w_\a= \xi^c (\g_c\l)_\a, \quad
\d \widehat w_\ah= \widehat\xi^c (\g_c\lh)_\ah.}

As in standard coset constructions, the 
${PSU(2,2|4)}\over{SO(4,1)\times SO(5)}$ coset
$G(x,\t,\widehat\t)$ is defined up to
right multiplication by a local $SO(4,1)\times SO(5)$ parameter 
$\O^{[ab]}(x,\t,\th)$ as
\eqn\gaugetr{\d G(x,\t,\th) = G(x,\t,\th) ~(\O^{[ab]}(x,\t,\th) T_{[ab]})}
where $T_{[ab]}$ are the 
$SO(4,1)\times SO(5)$ generators. Under these gauge transformations,
the pure spinors are defined to transform covariantly as
\eqn\puret{\d \l^\a = -\half \O^{[ab]} (\g_{[ab]}\l)^\a, \quad
\d w_\a = \half \O^{[ab]} (\g_{[ab]}w)_\a, }
$$\d \lh^\ah = -\half \O^{[ab]} (\g_{[ab]}\l)^\ah, \quad
\d \widehat w_\ah = \half \O^{[ab]} (\g_{[ab]}\widehat w)_\ah.$$

A convenient way to write the sigma model action in a manifestly
gauge-invariant manner is
\ref\Bersh{N. Berkovits, M. Bershadsky, T. Hauer, S. Zhukov and
B. Zwiebach, {\it Superstring theory on $AdS_2\times S^2$ as a coset
supermanifold}, Nucl. Phys. B567 (2000) 61, hep-th/9907200.}\purer
\eqn\sinv{S = {1\over \L} \int d^2 z
[ \half \eta_{AB} (J^A - {\cal A}^A)
(\bar J^B - \bar{\cal A}^B) }
$$+ {\cal B} + w_\a 
(\bar\p\l +\half \bar{\cal A}^{[ab]} \g_{[ab]}\l)^\a
+ \widehat w_\ah (\p\lh +\half
{\cal A}^{[ab]} \g_{[ab]}\lh)^\ah ]$$
$$ = 
 {1\over { \L}}\int d^2 z
 [\half\eta_{[ab][cd]} (J^{[ab]} - {\cal A}^{[ab]})
(\bar J^{[cd]} - \bar{\cal A}^{[cd]}) + \half \eta_{cd} J^c \bar J^d +
{1\over 4}\eta_{\a \bh}(J^\bh \bar J^\a + \bar J^\bh J^\a) $$
$$+
{1\over 2}\eta_{\a \bh}(J^\bh \bar J^\a - \bar J^\bh J^\a) 
+  
w_\a (\bar\p\l +\half \bar{\cal A}^{[ab]} \g_{[ab]}\l)^\a
+ \widehat w_\ah (\p\lh
+\half{\cal A}^{[ab]} \g_{[ab]}\lh)^\ah ], $$
where $\eta_{AB}$ is the $PSU(2,2|4)$ metric,
$\eta_{[ab][cd]} = \eta_{a[c}\eta_{d]b}$ when $a,b,c,d=0$ to 4,
$\eta_{[ab][cd]} = -\eta_{a[c}\eta_{d]b}$ when $a,b,c,d=5$ to 9,
$\eta_{cd}$ is the d=10 Minkowski metric, 
$\eta_{\a\bh} = (\g^{01234})_{\a\bh}$,
${\cal A}^{[ab]}$ and $\bar{\cal A}^{[ab]}$ are worldsheet
$SO(4,1)\times SO(5)$ gauge fields, 
and ${\cal B}$ is the Wess-Zumino term which in an $AdS_5\times S^5$
background takes the simple form \Bersh
\eqn\bads{{\cal B} ={1 \over 2} 
\eta_{\a\bh} (J^\bh \bar J^\a - \bar J^\bh J^\a).}

Since 
${\cal A}^{[ab]}$ and $\bar{\cal A}^{[ab]}$ satisfy auxiliary
equations of motion, they can be integrated out to obtain the action
\eqn\sinteg{S =  {1\over { \L}}\int d^2 z[ \half\eta_{cd} J^c \bar J^d +
\eta_{\a \bh}({3\over 4} J^\bh \bar J^\a -{1\over 4} \bar J^\bh J^\a) }
$$+
 w_\a (\bar\nabla \l)^\a + \widehat w_\ah (\nabla\lh)^\ah 
- \half \eta_{[ab][cd]} (w\g^{[ab]}\l)(\widehat w \g^{[cd]}\lh) ],$$
where
$(\bar\nabla \l)^\a = \bar\p\l^\a +\half \bar J^{[ab]} (\g_{[ab]} \l)^\a$ and
$(\nabla \lh)^\ah = \p\lh^\ah +\half J^{[ab]} (\g_{[ab]} \lh)^\ah.$
Using the Maurer-Cartan equations, the action of \sinteg\
can be shown to be
invariant under the BRST transformation generated by \adsv
\eqn\brsto{Q + \bar Q =  \int dz~ \eta_{\a\ah} \l^\a J^\ah + \int d\bar z
~\eta_{\a\ah}\lh^\ah \bar J^\a}
which transform the ${PSU(2,2|4)}\over{SO(4,1)\times SO(5)}$ 
coset and pure spinor ghosts as
\eqn\transc{\d G  = G (\e\l^\a T_\a + \e\lh^\ah T_\ah), \quad
\d w_\a = \e \eta_{\a\bh} J^\bh, \quad \d \widehat w_\ah = 
\e \eta_{\a\bh} \bar J^\bh, }
where $T_\a$ and $T_\ah$ are the 32 fermionic generators of $PSU(2,2|4)$
and $\e$ is a constant Grassmann parameter.

This BRST invariance, together with $PSU(2,2|4)$ invariance,
fixes the relative coefficients of the terms in the sigma model action
of \sinteg. So, naively, the $AdS_5\times S^5$ radius $r$ can only appear in
the action through the coupling constant $\L = \a'/r^2$.
However, if one allows the $PSU(2,2|4)$ algebra to be deformed as
the value of $r$ is changed, the $r$ dependence of the action can
be more complicated and the form of the action can be modified. 
For example, in the flat space limit where $r\to \infty$,
the $PSU(2,2|4)$ algebra is deformed to the N=2 d=10 super-Poincar\'e algebra.
As will now be discussed, this modifies the sigma model action
of \sinteg\ to a quadratic action.

\subsec{Flat space limit}

Although the naive limit as $r\to \infty$ is obtained by simply taking
$\L\to 0$ in the sigma model action of \sinteg, this limit would preserve
$PSU(2,2|4)$ invariance instead of the desired N=2 d=10 super-Poincar\'e
invariance of flat Minkowski superspace. To obtain the correct flat space
limit, one needs to rescale the $PSU(2,2|4)$ structure constants such
that when $r\to\infty$, the $PSU(2,2|4)$ algebra is deformed into the
N=2 d=10 super-Poincar\'e algebra. 

The non-vanishing $PSU(2,2|4)$ structure constants $f_{AB}^C$ are 
\eqn\stps{
f_{\a\b}^{{c}} = \g^{{c}}_{\a\b},\quad
f_{\ah\bh}^{{c}} = \g^{{c}}_{\a\b},}
$$f_{\a c}^\bh
=-   \g_{c\a\b}
\d^{\b\bh},\quad
f_{\ah c}^\b =
- 
\g_{c\ah\bh} \d^{\b\bh},$$
$$
f_{\a \bh}^{[ef]}= \pm 
(\g^{ef})_\a{}^\g \d_{\g\bh},\quad
f_{c d}^{[ef]}= \pm   \d_c^{[e} \d_d^{f]},$$
$$f_{[{cd}][{ef}]}^{[{gh}]}=
\eta_{{ce}}\d_{{d}}^{[{g}}
\d_{{f}}^{{h}]}
-\eta_{{cf}}\d_{{d}}^{[{g}}
\d_{{e}}^{{h}]}
+\eta_{{df}}\d_{{c}}^{[{g}}
\d_{{e}}^{{h}]}
-\eta_{{de}}\d_{{c}}^{[{g}}
\d_{{f}}^{{h}]},$$
$$f_{[{cd}] e}^{f} =
\eta_{{e}
{[c}} \d_{d]}^{{f}},\quad
f_{[{cd}] \a}^{\b} =
\half(\g_{{cd}})_\a{}^\b,\quad
f_{[{cd}] \ah}^{\bh} =
\half(\g_{{cd}})_\ah{}^\bh,$$
where the $+$ sign in the third line is if $(c,d,e,f)= 0$ to 4,
and the $-$ sign is if $(c,d,e,f) = 5$ to 9.

To deform these structure constants to the super-Poincar\'e structure
constants in the $r\to\infty$ limit, one should rescale \stps\ such
that 
\eqn\stpa{
f_{\a\b}^{{c}} = \g^{{c}}_{\a\b},\quad
f_{\ah\bh}^{{c}} = \g^{{c}}_{\a\b},}
$$f_{\a c}^\bh
=-  r^{-1} \g_{c\a\b}
\d^{\b\bh},\quad
f_{\ah c}^\b =
- r^{-1}
\g_{c\ah\bh} \d^{\b\bh},$$
$$
f_{\a \bh}^{[ef]}= \pm r^{-2}
(\g^{ef})_\a{}^\g \d_{\g\bh},\quad
f_{c d}^{[ef]}= \pm  r^{-2}  \d_c^{[e} \d_d^{f]},$$
$$f_{[{cd}][{ef}]}^{[{gh}]}=
\eta_{{ce}}\d_{{d}}^{[{g}}
\d_{{f}}^{{h}]}
-\eta_{{cf}}\d_{{d}}^{[{g}}
\d_{{e}}^{{h}]}
+\eta_{{df}}\d_{{c}}^{[{g}}
\d_{{e}}^{{h}]}
-\eta_{{de}}\d_{{c}}^{[{g}}
\d_{{f}}^{{h}]}$$
$$f_{[{cd}] e}^{f} =
\eta_{{e}
{[c}} \d_{d]}^{{f}},\quad
f_{[{cd}] \a}^{\b} =
\half(\g_{{cd}})_\a{}^\b,\quad
f_{[{cd}] \ah}^{\bh} =
\half(\g_{{cd}})_\ah{}^\bh.$$

The metric $g_{AB}$ should satisfy the
property that $f_{AB}^C ~g_{CD}$ is
graded-antisymmetric under permutations of $[ABD]$,
so the rescaling of \stpa\ implies one
should also rescale $g_{\a\bh} = \eta_{\a\bh}$ and $g_{[ab][cd]} =
\eta_{[ab][cd]}$ to
\eqn\rescone{g_{\a\bh} = r \eta_{\a\bh}, \quad
g_{[ab][cd]} = r^2 \eta_{[ab][cd]}.}

Since the structure constants $f_{AB}^C$ are proportional to the
superspace torsions $T_{AB}{}^C$, the rescaling of \stpa\ implies that
\eqn\ratiosa
{{{T_{\a \b}{}^b \eta_{ab}}\over{T_{\a a}{}^\bh \eta_{\b\bh}}} = r.}
If $T_{\a\b}{}^b$ is fixed to satisfy
$T_{\a\b}{}^b = \g_{\a\b}^b$, \ratiosa\ implies
that
$T_{\a c}{}^\bh = r^{-1} \g_{c\a\b}\eta^{\b\bh}$, which is the correct
$r$ dependence since the $AdS$ curvature 
$R_{ab\a}{}^\b$ goes like $1/r^2$, and Bianchi identities imply that
$R_{ab\a}{}^\b $ is proportional to $T_{a\a}{}^\g T_{b\g}{}^\b$.

Since $g_{\a\bh} = r \eta_{\a\bh}$ blows up when $r\to\infty$, it
is convenient to write the second-order kinetic term for the fermions 
in \sinteg\ in the first-order form as
\eqn\fermtwo{{1\over {\L}}\int d^2 z ~ 
r\eta_{\a \bh}({3\over 4} J^\bh \bar J^\a -{1\over 4} \bar J^\bh J^\a) }
$$= 
{1\over {\L}}\int d^2 z ~
r \eta_{\a \bh} (\half J^\bh \bar J^\a + {1\over 4} J^\bh \wedge J^\a) $$
$$ = 
{1\over {\L}}\int d^2 z 
[\bar J^\a d_\a + J^\ah \widehat d_\ah + 2 r^{-1} \eta^{\a\bh} 
d_\a \widehat d_\bh + {1\over 4}
r\eta_{\a\bh}\int d\s_3~ d(J^\bh \wedge J^\a)] $$
$$ = 
{1\over {\L}}\int d^2 z 
[\bar J^\a d_\a + J^\ah \widehat d_\ah + 2 r^{-1} \eta^{\a\bh} 
d_\a \widehat d_\bh 
+ {1\over 4}\int d\s_3 (\g_{c\a\b} J^c \wedge J^\a \wedge J^\bh
-\g_{c\ah\bh} J^c \wedge J^\ah\wedge J^\bh)] $$
where $d_\a$ and $\widehat d_\ah$ are auxiliary variables and the two-form
$J^\bh\wedge J^\a \equiv J^\bh \bar J^\a - \bar J^\bh J^\a$ has been
written as the integral of a Wess-Zumino-Witten three-form using the 
Maurer-Cartan equations
\eqn\mce{  dJ^\bh = f_{c\a}^\bh J^c \wedge J^\a =
r^{-1}\g_{c\a\b} \eta^{\b\bh}
J^c \wedge J^\a,}
$$dJ^\b = f_{c\ah}^\b J^c \wedge J^\ah =  r^{-1}\g_{c\ah\bh} \eta^{\b\bh}
J^c \wedge J^\ah.$$
Furthermore, the BRST operator $Q+\bar Q$ of \brsto\ can be written as
\eqn\brsttwo{Q + \bar Q = \int dz \l^\a d_\a + \int d\bar z \lh^\ah
\widehat d_\ah} using the auxiliary equations
of motion for $d_\a$ and $\widehat d_\ah$.

When $r= \infty$, the left-invariant currents $(J^c, J^\a, J^\bh, J^{[ab]})$
simplify to
\eqn\jsimp{J^c = \Pi^c = \p x^c + \t \g^c \p\t + \th \g^c \p\th,\quad
J^\a=\p\t^\a,\quad J^\bh = \p\th^\bh,\quad J^{[ab]}=0.}
So the action of \sinteg\ reduces to
\eqn\sflat{ S = {1\over \L} \int d^2 z
[\half \eta_{cd}\Pi^c \bar\Pi^d - d_\a \bar\p\t^\a - \widehat d_\ah
\p\th^\ah   + w_\a\bar\p\l^\a + \widehat w_\ah\p\lh^\ah}
$$
+{1\over 4} \int d\s_3 (\g_{c\a\b} \Pi^c \wedge \p\t^\a \wedge \p\t^\b
-\g_{c\ah\bh} \Pi^c \wedge \p\th^\ah\wedge \p\th^\bh)],  $$
which is the worldsheet action in a flat background using the pure
spinor formalism. By defining 
\eqn\defp{p_\a = d_\a + ..., \quad \widehat p_\ah = \widehat d_\ah + ...}
where $...$ are functions of $(x,\t,\th)$, this action can be written
in quadratic form as \purer
\eqn\squad{ S = {1\over \L} \int d^2 z
[\half \eta_{cd}\p x^c \bar\p x^d - p_\a \bar\p\t^\a - \widehat p_\ah
\p\th^\ah   + w_\a\bar\p\l^\a + \widehat w_\ah\p\lh^\ah].}

\newsec{New Limit of Sigma Model}

In the previous section, we constructed the flat space limit of
the $AdS_5\times S^5$ sigma model in which $T_{c\a}{}^\bh \to 0$
and $T_{\a\b}{}^c = \g_{\a\b}^c$. In this section, we shall consider
a different limit of the model in which 
$T_{\a\b}{}^c \to 0$
and $T_{c\a}{}^\bh = \g_{c\a\b} \eta^{\b\bh}$. If one defines $r$ as
in \ratiosa, this formally corresponds
to the limit $r\to 0$ of the $AdS_5\times S^5$ background. However,
since supergravity backgrounds are usually defined such that
$T_{\a\b}{}^c = \g_{\a\b}^c$ \howew, this limit cannot be identified with
a conventional supergravity background. 

\subsec{$T_{\a\b}{}^c \to 0$ limit}

To construct the sigma model in this new limit, one needs to rescale
the $PSU(2,2|4)$ structure constants of \stps\ as
\eqn\stpt{
f_{\a\b}^{{c}} = r \g^{{c}}_{\a\b},\quad
f_{\ah\bh}^{{c}} = r\g^{{c}}_{\a\b},}
$$f_{\a c}^\bh
=-  \g_{c\a\b}
\d^{\b\bh},\quad
f_{\ah c}^\b =
-
\g_{c\ah\bh} \d^{\b\bh},$$
$$
f_{\a \bh}^{[ef]}= \pm r 
(\g^{ef})_\a{}^\g \d_{\g\bh},\quad
f_{c d}^{[ef]}= \pm    \d_c^{[e} \d_d^{f]},$$
$$f_{[{cd}][{ef}]}^{[{gh}]}=
\eta_{{ce}}\d_{{d}}^{[{g}}
\d_{{f}}^{{h}]}
-\eta_{{cf}}\d_{{d}}^{[{g}}
\d_{{e}}^{{h}]}
+\eta_{{df}}\d_{{c}}^{[{g}}
\d_{{e}}^{{h}]}
-\eta_{{de}}\d_{{c}}^{[{g}}
\d_{{f}}^{{h}]}$$
$$f_{[{cd}] e}^{f} =
\eta_{{e}
{[c}} \d_{d]}^{{f}},\quad
f_{[{cd}] \a}^{\b} =
\half(\g_{{cd}})_\a{}^\b,\quad
f_{[{cd}] \ah}^{\bh} =
\half(\g_{{cd}})_\ah{}^\bh.$$
Furthermore, to preserve the graded-antisymmetry of 
$f_{AB}^C~ g_{CD}$ under permutation of $[ABD]$,
one needs to also rescale $g_{ab} = \eta_{ab}$ and $g_{[ab][cd]} =
\eta_{[ab][cd]}$ to
\eqn\resctwo{g_{ab} = r^{-1} \eta_{ab}, \quad
g_{[ab][cd]} = r^{-1} \eta_{[ab][cd]}.}

When $r\to 0$, the structure constants $f_{\a\b}^A\to 0$ which implies
that the 32 fermionic isometries become abelian. In this limit,
the ${PSU(2,2|4)}\over{SO(4,1)\times SO(5)}$ coset 
$G$ splits into a bosonic coset $H_{r'}^{r}$ for $r,r'=1$ to 4
which parameterizes $AdS_5={{SU(2,2)}\over {SO(4,1)}}$, a bosonic coset
$\Ht_{j'}^{j}$ for $j,j'=1$ to 4
which parameterizes $S^5={{SU(4)}\over {SO(5)}}$, and two fermionic matrices
$\t^{rj}$ and $\tb_{jr}$ for $r,j=1$ to 4. 
The index $r=1$ to 4
labels a fundamental representation of the global $SU(2,2)$, 
and the index $j=1$ to 4 
labels a fundamental representation of the global $SU(4)$. 
Furthermore, the index $r'=1$ to 4 labels a spinor representation of
the local $SO(4,1)$, and the index $j'=1$ to 4 labels a 
spinor representation of
of the local $SO(5)$. Note that $r'$ indices can be raised and lowered
with an antisymmetric $SO(4,1)$-invariant tensor $\e^{r' s'}$,
and $j'$ indices can be raised and lowered
with an antisymmetric $SO(5)$-invariant tensor $\e^{j' k'}$.
Under the 32 global fermionic isometries, 
\eqn\transt{\d\t^{rj} = \a^{rj}, \quad
\d\tb_{jr} = \bar\a_{jr}, \quad \d H^r_{r'} =0,\quad \d\Ht^j_{j'}=0, }
where $\a^{rj}$ and $\a_{jr}$ are constant Grassmann parameters. 

Since $g_{ab} = r^{-1}\eta_{ab}$ blows up when $r\to 0$, it is convenient
to write the second-order kinetic term for the bosons in the first-order
form as
\eqn\first{
{1\over {2\L}}\int d^2 z
[ r^{-1}\eta_{[ab][cd]} (J^{[ab]} - {\cal A}^{[ab]})
(\bar J^{[cd]} - \bar{\cal A}^{[cd]}) + r^{-1} \eta_{cd} J^c \bar J^d 
]}
$$=
{1\over \L}\int d^2 z
[(J^{[ab]} - {\cal A}^{[ab]} ) \bar P_{[ab]}
 +(\bar J^{[ab]} - \bar {\cal A}^{[ab]} ) P_{[ab]} 
+  J^c \bar P_c  + \bar J^c P^c $$
$$  + 2 r
(\eta^{[ab][cd]} P_{[ab]}\bar P_{[cd]} + \eta^{cd} P_c \bar P_d)]$$
where $[P_{[ab]}, \bar P_{[ab]}, P_c, \bar P_c]$ are auxiliary fields.
So the $AdS_5\times S^5$ sigma model action of \sinv\ reduces in this
limit $r\to 0$ to
\eqn\snew{S = {1\over \L} \int d^2 z
[(J^{[ab]} - {\cal A}^{[ab]} ) \bar P_{[ab]}
 +(\bar J^{[ab]} - \bar {\cal A}^{[ab]} ) P_{[ab]} 
+  J^c \bar P_c  + \bar J^c P^c }
$$+ {1\over 4}\eta_{\a\bh}(J^\bh \bar J^\a + \bar J^\bh J^\a) + 
{\cal B} + w_\a 
(\bar\p\l +\half \bar{\cal A}^{[ab]} \g_{[ab]}\l)^\a
+ \widehat w_\ah (\p\lh
+\half{\cal A}^{[ab]} \g_{[ab]}\lh)^\ah ]$$
where ${\cal B}$ is the Wess-Zumino-Witten term of \bads.
Since $\int d^2 z {\cal B}= {1\over 2}
\int d^2 z \int d\s_3 (\g_{c\a\b} J^c \wedge J^\a \wedge J^\b
-\g_{c\ah\bh} J^c \wedge J^\ah\wedge J^\bh), $
the Wess-Zumino-Witten term 
can be eliminated from the action by shifting $P_c$ and $\bar P_c$.

Furthermore, when $r\to 0$, the currents $J^c$ and $J^{[cd]}$
simplify to 
\eqn\simtwo{ J^c = (H^{-1} \p H)_{r'}^{s'} (\s^c)^{r'}_{s'}, \quad
J^{[cd]} = (H^{-1} \p H)_{r'}^{s'} (\s^{[cd]})^{r'}_{s'}\quad
{\rm when ~~ }c,d=0~~{\rm to}~~4,}
$$J^c = (\Ht^{-1} \p \Ht)_{j'}^{k'} (\s^c)^{j'}_{k'}, \quad
J^{[cd]} = (\Ht^{-1} \p \Ht)_{j'}^{k'} (\s^{[cd]})^{j'}_{k'}\quad
{\rm when ~~ }c,d=5~~{\rm to}~~9,$$
where $\s^c$ and $\s^{[cd]}$ are $4\times 4$ Pauli matrices which
generate an $SU(2,2)$ algebra when $c=0$ to 4, and generate an
$SU(4)$ algebra when $c=5$ to 9. Expressing the $SO(9,1)$ spinors 
$J^\a$ and $J^\ah$ in terms of $SO(4,1)\times SO(5)$ spinors as
$J^\a = J^{r'j'}$ and $J^\ah = \widehat J^{r'j'}$, one finds
that when $r\to 0$,
$J^{r'j'}$ and $\widehat J^{r'j'}$ simplify to 
\eqn\simpcurr{J^{r' j'} = (H^{-1})_{r}^{r'} (\Ht^{-1})_{j}^{j'} \p\t^{r j} +
\e^{r's'}\e^{j'k'}H_{s'}^{r} \Ht_{k'}^j \p\bar\t_{j r}, }
$$\widehat J^{r' j'} = (H^{-1})_{r}^{r'} (\Ht^{-1})_{j}^{j'} \p\t^{r j} -
\e^{r's'}\e^{j'k'}H_{s'}^{r} \Ht_{k'}^j \p\bar\t_{j r}.$$
Plugging these currents into \snew, one finds that the action
simplifies to 
\eqn\snewt{S = {1\over \L} \int d^2 z
[(J^{[ab]} - {\cal A}^{[ab]} ) \bar P_{[ab]}
 +(\bar J^{[ab]} - \bar {\cal A}^{[ab]} ) P_{[ab]} 
+  J^c \bar P_c  + \bar J^c P^c }
$$+  \p \tb_{j r}\p\t^{r j} 
+ w_\a 
(\bar\p\l +\half \bar{\cal A}^{[ab]} \g_{[ab]}\l)^\a
+ \widehat w_\ah (\p\lh
+\half {\cal A}^{[ab]} \g_{[ab]}\lh)^\ah ].$$

\subsec{Twistor-like variables}

The final step in simplifying this action is to express the pure
spinors in $SO(4,1)\times SO(5)$ notation as
$\l^\a = \l^{r'j'}$ and $\lh^\ah = \lh^{r'j'}$ and to define
the new variables $Z^{r j}$ and $\bar Z_{jr}$ as
\eqn\twistone{ Z^{rj} = H^r_{r'} \Ht^j_{j'} \l^{r'j'},\quad
\bar Z_{jr} = (H^{-1})_r^{r'} (\Ht^{-1})_j^{j'} \lh_{j'r'}}
where $\lh_{j'r'} = \e_{j'k'}\e_{r's'} \lh^{s'k'}$. 
Note that $Z^{r j}$ and $\bar Z_{jr}$ are twistor-like
variables since they transform covariantly under the global
$SU(2,2)\times SU(4)$ isometries and since they are constructed
out of the pure spinors and the ten $x$'s
parameterized by the cosets $H$ and $\Ht$. 
Similarly, one can define the conjugate twistor-like variables
$Y_{jr}$ and $\bar Y^{rj}$ as
\eqn\twisttwo{Y_{jr} = (H^{-1})_r^{r'} (\Ht^{-1})_j^{j'} w_{j'r'},\quad
\bar Y^{rj} = H^r_{r'} \Ht^j_{j'} \widehat w^{r'j'}}
where $w_\a = w_{j'r'}$ and $\widehat w_\ah = \e_{j'k'}\e_{r's'}
\widehat w^{s'k'}$
are the original conjugate pure spinor variables written in
$SO(4,1)\times SO(5)$ notation.

Using
\eqn\wsimp{Y_{jr}\bar\p Z^{jr} = w_\a\bar\p \l^\a
+(H^{-1}\bar\p H)^{r'}_{s'} w_{j'r'} \l^{s'j'} +
(\Ht^{-1}\bar\p \Ht)^{j'}_{k'}w_{j' r'}\l^{r'k'},}
one finds that  
\eqn\simpt{w_\a \bar\p \l^\a = Y_{jr}\bar\p Z^{rj}
- (w \s_c \l) \bar J^c -\half (w\s_{[cd]}\l) \bar J^{[cd]}}
where $(w\s_c\l)= w_{j'r'} (\s_c)_{s'}^{r'}\l^{s'j'}$ and
$(w\s_{[cd]}\l)= w_{j'r'} (\s_{[cd]})_{s'}^{r'}\l^{s'j'}$ for $c=0$ to 4,
and 
$(w\s_c\l)= w_{j'r'} (\s_c)_{k'}^{j'}\l^{r'k'}$ and
$(w\s_{[cd]}\l)= w_{j'r'} (\s_{[cd]})_{k'}^{j'}\l^{r'k'}$ for $c=5$ to 9.
Similarly, 
\eqn\simpu{\widehat w_\ah \p \lh^\ah = \bar Y^{r j}\p \bar Z_{jr}
- (\widehat w \s_c \lh) J^c -\half (\widehat w\s_{[cd]}\lh)  J^{[cd]}.}
So after defining 
\eqn\shiftp{P'^c = P^c - (w\s^c \l), \quad
\Pbt^c = \bar  P^c -(\widehat w\s^c \lh), }
$$P'^{[cd]} = P^{[cd]} -\half( w\s^{[cd]} \l), \quad
\Pbt^{[cd]} =\bar  P^{[cd]} -\half(\widehat w\s^{[cd]} \lh), $$
one can write the action of \snewt\ as 
\eqn\actf{S=  {1\over \L} \int d^2 z
[(J^{[ab]} - {\cal A}^{[ab]} ) \Pbt_{[ab]}
 +(\bar J^{[ab]} - \bar {\cal A}^{[ab]} ) P'_{[ab]} 
+  J^c \Pbt_c  + \bar J^c P'^c }
$$+  \p \tb_{j r}\p\t^{r j} + Y_{jr}\bar\p Z^{rj} + \bar Y^{rj}\p \bar
Z_{jr}].$$

The shift of \shiftp\ implies that under the gauge transformation
$\d w_\a = 
 \xi^c (\g_c\l)_\a$ and $\d \widehat w_\ah= \widehat\xi^c (\g_c\lh)_\ah$
of \gaugew, $P'_c$ and $\bar P'_c$ must transform as 
\eqn\transp{\d P'_c = \xi^c \e_{r's'}\e_{j'k'}\l^{r'j}\l^{s'k'}=
\xi^c(\l\g^{01234}\l), }
$$\d \bar P'_c = \widehat\xi^c \e^{r's'}\e^{j'k'}\lh_{r'j}\lh_{s'k'}=
\widehat\xi^c(\lh\g^{01234}\lh).$$
So assuming that $(\l\g^{01234}\l)$ and 
$(\lh\g^{01234}\lh)$ are non-zero, one can use this invariance to
gauge-fix $P'^c= \bar P'^c=0$.
Furthermore, integrating out ${\cal A}^{[ab]}$ and $\bar{\cal A}^{[ab]}$ 
implies that $P'^{[ab]}= \bar P'^{[ab]}=0$. 

So finally, one can write the action in quadratic form as
\eqn\actfi{S = {1\over \L} \int d^2 z
[ \p \tb_{j r}\bar\p\t^{r j} + Y_{jr}\bar\p Z^{rj} + \bar Y^{rj}\p \bar
Z_{jr}].}
Instead of the original action containing ten $x$'s and 22 left and 
right-moving pure spinors, \actfi\ contains 16 left-moving and
16 right-moving unconstrained bosonic spinors. So the second-order action
for $x$ has been converted into a first-order action for ten left and
right-moving bosons which effectively removes the constraint on the
pure spinors.
The removal of the pure spinor constraint is related to the fact that
$T_{\a\b}{}^c=0$ in this background.
Since the BRST operator acts as $Q= \l^\a \nabla_\a$, $Q^2= \l^\a \l^\b
\{\nabla_\a,\nabla_\b\} =\l^\a\l^\b
T_{\a\b}{}^A \nabla_A$. When $T_{\a\b}{}^c=
\g_{\a\b}^c$, the pure spinor constaint $\l\g^c\l=0$ is required for
$Q$ to be nilpotent. However, when $T_{\a\b}{}^c=0$, the nilpotence
of $Q$ does not require 
$\l^\a$ to satisfy the pure spinor constraint.

\subsec{N=2 worldsheet supersymmetry}

In terms of the variables $(\t^{rj}, \tb_{jr}, Z^{rj}, \bar Z_{jr}, Y_{jr},
\bar Y^{rj})$, the BRST transformations are
\eqn\brsta{\d\t^{rj} = \e Z^{rj}, \quad \d\tb_{jr} = \e \bar Z_{jr},
\quad \d Y_{jr} = \e \p\tb_{rj}, \quad \d \bar Y^{rj}= \e\bar\p\t^{rj},}
which are generated by $Q+\bar Q$ where
\eqn\brstb{Q = \int dz Z^{rj}\p\tb_{jr}, \quad \bar Q= \int d\bar z \bar Z_{jr}
\bar\p\t^{rj}.}
Unlike in a flat background where it is difficult to construct 
$b$ and $\bar b$ ghosts satisfying $\{Q,b\}=T$ and $\{\bar Q, \bar b\}=
\bar T$, it is easy to construct $b$ and $\bar b$ ghosts in this background
as 
\eqn\bgh{b =  Y_{jr}\p\t^{rj}, \quad \bar b= \bar Y^{rj} \bar\p\tb_{jr},}
where 
\eqn\tdef{T =  \p\t^{rj}\p\tb_{jr}+ Y_{jr}\p Z^{rj}, \quad 
\bar T=
 \bar\p\t^{rj}\bar\p\tb_{jr}+\bar Y^{rj}\bar\p \bar Z_{jr}.}

Since $Y_{jr}$ and $\bar Y^{jr}$ have conformal weight $(1,0)$ and
$(0,1)$, 
the action of \actfi\ has A-twisted N=(2,2) supersymmetry
and can be interpreted as a topological A-model. This topological
A-model can be expressed in N=(2,2) superspace by combining the
component fields into the chiral and antichiral superfields
\eqn\superd{\T^{rj} = \t^{rj} + \k_+ Z^{rj} + \k_- \bar Y^{rj} + \k_+\k_-
f^{rj},}
$$\Tb_{jr} = \tb_{jr} + \kb_+ Y_{jr} + \kb_- \bar Z_{jr} + \kb_+\kb_- \bar
f_{jr},$$
where $(\k_+,\kb_+)$ and $(\k_-,\kb_-)$ are the left and right-moving
N=(2,2) Grassmann parameters, and $(f^{rj},\bar f_{jr})$ are auxiliary
fields.

In terms of $\T^{rj}$ and $\Tb_{jr}$, the action of \actfi\ is
\eqn\actsuper{S = 
{1\over \L} \int d^2 z\int d^4\k \Tb_{jr}\T^{rj},} 
and the global bosonic isometries act as 
\eqn\actf{\d\T^{r j} = i \L^r_s \T^{s j} + i \T^{r k} \O_k^j + i \S \T^{r j},
\quad
\d\Tb_{j r} =- i \Tb_{j s}\L^s_r - i \O_j^k\Tb_{k r} - i \S \Tb_{j r},}
where $(\L^r_s, \Omega_j^k, \S)$
are constant parameters satisfying
$\L^r_r = \O^j_j=0$. 
Note that in addition to the $SU(2,2)\times SU(4)$ bosonic isometries,
there is an additional ``bonus''
$U(1)$ symmetry parameterized by $\S$.
Under the fermionic isometries of \transt, the superfields transform as
\eqn\actg{\d\T^{r j} = \a^{rj},
\quad
\d\Tb_{j r} =\bar\a_{jr}.}

\newsec{Non-Linear Topological A-Model}

To compute the physical states of the linear topological A-model of
\actsuper, it will be useful to define a non-linear topological
A-model which reduces to the linear model of \actsuper\ in
a certain large-radius limit. In the non-linear model, the $SU(2,2)\times
SU(4)\times U(1)$ bosonic isometries will combine with the 32
fermionic isometries to form an $SU(2,2|4)$ supergroup. Since this
supergroup includes the $PSU(2,2|4)$ isometries of the $AdS_5\times S^5$
background, it is tempting to try to identify this non-linear topological
A-model at large but finite radius with the $AdS_5\times S^5$
sigma model at small but non-zero $T_{\a\b}{}^c$.
However, this identification does not seem possible since when
$T_{\a\b}{}^c$ is non-zero, the $AdS_5\times S^5$ sigma model contains
a Wess-Zumino-Witten term which is antisymmetric under exchange of 
$z$ and $\bar z$ and which breaks $SU(2,2|4)$ down to $PSU(2,2|4)$.
On the other hand, the non-linear topological A-model is symmetric
under exchange of $z$ and $\bar z$ and preserves $SU(2,2|4)$
invariance. So it appears that the $AdS_5\times S^5$ sigma model and
the non-linear topological A-model can only be identified in the limit
where $T_{\a\b}{}^c=0$ in the $AdS_5\times S^5$ model and where
the radius is infinite in the non-linear model.

\subsec{Superspace action}

Although the non-linear topological A-model has both N=(2,2)
worldsheet supersymmetry and $SU(2,2|4)$ invariance, both these symmetries can
not be simultaneously made manifest.
The worldsheet supersymmetry
can be made manifest by expressing the non-linear action in superspace as
\eqn\actftwo{S = {1\over \L} \int d^2 z d^4 \k
[ \Tb_{r j}\T^{j r} - {1\over {2R^2}} \Tb_{r j}\T^{j s}\Tb_{s k}\T^{kr}
+ {1\over {3R^4}}
\Tb_{r j}\T^{j s}\Tb_{s k}\T^{k t} \Tb_{t l}\T^{l r} + ... ] }
$$ = {R^2\over \L} \int d^2 z d^4 \k ~ Tr [ \log (1 + {1\over {R^2}}
\Tb \T) ]$$
where $\T_{rj}$ and $\Tb_{jr}$ are the same superfields as in
\superd, and 
$R$ is the radius of this model which is unrelated to the 
$AdS_5\times S^5$ radius $r$.
In the limit $R\to \infty$, this non-linear model reduces to 
the linear topological A-model of \actsuper.
The non-linear action of \actftwo\ is invariant under the same
$SU(2,2)\times SU(4)\times U(1)$ transformations as \actf, but
the fermionic isometries of \actg\ are modified to 
\eqn\acth{\d\T^{r j} = \a^{r j} + {1\over R^2}
\T^{r k}\bar\a_{k s}\T^{s j}, \quad
\d\Tb_{j r} = \bar\a_{j r}  +{1\over R^2} \Tb_{j s}\a^{s k} \Tb_{k r},}
which close with the bosonic isometries into the $SU(2,2|4)$ supergroup.

\subsec{Coset action}

These $SU(2,2|4)$ isometries can be made manifest by rescaling
$\T^{rj}\to R\T^{rj}$ and
$\Tb_{jr}\to R\Tb_{jr}$ and writing the non-linear action in terms
of the component fields $(\t^{rj}, \tb_{jr}, Z^{rj},\bar Z_{jr},
Y_{jr}, \bar Y^{rj})$ using a coset space construction. The coset
$G$ will be defined to take values in ${PSU(2,2|4)}\over{SU(2,2)\times 
SU(4)}$,
and since the coset has only fermionic elements, $G$ can be 
gauged to the form
\eqn\gaugeg{G_j^k = \d_j^k, \quad G_s^r = \d_s^r, \quad
G^{rj}= \t^{rj}, \quad G_{jr} = \tb_{jr}.}

In terms of the left-invariant currents $J^A = (G^{-1} \p G)^A$ and
$\bar J^A = (G^{-1}\bar\p G)^A$ where $A$ is an $SU(2,2|4)$ index,
the action is
\eqn\coset{S = 
 {{R^2}\over { \L}}\int d^2 z
 [(\bar J-\bar {\cal A})_s^r (J-{\cal A})_r^s
 -(\bar J-\bar {\cal A})_j^k (J-{\cal A})_k^j  }
$$+ \bar J_{jr} J^{rj} +
Y_{jr} (\bar\p Z+ \bar{\cal A}Z)^{rj}
+ \bar Y^{rj} (\p \bar Z -{\cal A}\bar Z)_{jr} ] $$
\eqn\cosetp{= 
 {{R^2}\over { \L}}\int d^2 z
[\bar J_{jr} J^{rj} +
Y_{jr} \bar\nabla Z^{rj}
+ \bar Y^{rj} \nabla \bar Z_{jr}  +
Y_{jr}Z^{rk}\bar Z_{ks}\bar Y^{sj} -  
Z^{rj}Y_{js}\bar Y^{sk}\bar Z_{kr}]}
where $({\cal A}^A, \bar{\cal A}^A)$ are $SU(2,2)\times SU(4)$
gauge fields,
$\bar\nabla Z^{jr} = \bar\p Z^{jr} + \bar J^r_s Z^{js}
+ \bar J^j_k Z^{kr}$, and
$\nabla \bar Z_{rj} = \p \bar Z_{rj} - J^s_r \bar Z_{sj}
- J^k_j \bar Z_{rk}$.
Note that 
\eqn\notone{\bar J_{jr} J^{rj} - J_{jr}\bar J^{rj} =
\p \bar J_{U(1)} - \bar \p J_{U(1)}}
is a total derivative where $J_{U(1)}$ is the ``bonus'' $U(1)$ current,
so the term $\int d^2 z \bar J_{jr} J^{rj}$
is symmetric under exchange of $z$ and $\bar z$.

Although $SU(2,2|4)$ invariance is manifest in the action of \coset,
N=(2,2) worldsheet supersymmetry is not manifest. Nevertheless, one
can easily construct the twisted N=(2,2) worldsheet supersymmetry
generators as
\eqn\gens{Q = \int dz Z^{rj} J_{jr},\quad \bar Q = \int d\bar z
\bar Z_{jr} \bar J^{rj},\quad
b = Y_{jr} J^{rj},\quad \bar b = \bar Y^{rj} \bar J_{jr}.}
After 
parameterizing
$G$ as in \gaugeg, the action of \cosetp\ coincides with the superspace
action of \actftwo\ after integrating out the auxiliary fields
$f^{rj}$ and $\bar f_{jr}$. 

\subsec{One-loop conformal invariance}

To show that the non-linear topological A-model has no one-loop
conformal anomaly, one can either use the superspace version of the
action of \actftwo\ and compute $\log ~\det (\p \bar\p K)$ where $K$ is the 
Kahler potential, or one can use the coset version of the action 
of \cosetp\ and compute the anomaly with the background field method of
\Bersh\
and \vall.
Absence of this anomaly is necessary for the topological twisting to
be consistent at the quantum level.

Using the superspace action of \actftwo, $K= Tr~ \log(1+\Tb\T)$ implies that
\eqn\superm{\p_{ks}\bar\p^{rj}K =
\p_{ks}[ \T^{rl} [(1+\Tb\T)^{-1}]^j_l ]}
$$= \d^r_s 
[(1+\Tb\T)^{-1}]^j_k  - 
\T^{rl} 
[(1+\Tb\T)^{-1}]^m_l \Tb_{ms}
[(1+\Tb\T)^{-1}]^j_k $$
$$= 
[(1+\T\Tb)^{-1}]^r_s 
[(1+\Tb\T)^{-1}]^j_k .$$
So there is no conformal anomaly since
\eqn\supert{\log~\det (
\p_{ks}\bar\p^{rj}K) = \log~\det 
[(1+\T\Tb)^{-1}] + \log~\det
[(1+\Tb\T)^{-1}]}
$$= -Tr ~\log (1+\T\Tb) -
Tr ~\log (1+\Tb\T) 
= - Tr [ \sum_{n=1}^\infty {{(-1)^{n+1}}\over n} (\T\Tb)^n
+ \sum_{n=1}^\infty {{(-1)^{n+1}}\over n} (\Tb\T)^n ] = 0$$
where we have used that $Tr [(\T\Tb)^n] = -Tr [(\Tb\T)^n]$ for $n>0$.

Using the background field method for the coset action of \cosetp, 
the matter sector
of $\int d^2 z \bar J_{jr} J^{rj}$ contributes no conformal anomaly since,
when $G/H$ is a symmetric space, the $G/H$ coset model has the same
conformal anomaly as the principal chiral model based on $G$ \Bersh.
In this case, $PSU(2,2|4)/(SU(2,2)\times SU(4))$ is a symmetric space,
and the principal chiral model based on $PSU(2,2|4)$ has no conformal
anomaly \ref\witber{N. Berkovits, C. Vafa and E. Witten,
{\it Conformal field theory of AdS background with Ramond-Ramond flux},
JHEP 9903 (1999) 018, hep-th/9902098.}.

Furthermore, the ghost sector of \cosetp\ contributes no conformal anomaly
because of a cancellation between the $Y_{jr}\bar\nabla Z^{rj} +
\bar Y^{rj}\nabla \bar Z_{jr}$ contribution and the 
$Y_{jr}Z^{rk}\bar Z_{ks}\bar Y^{sj} -  
Z^{rj}Y_{js}\bar Y^{sk}\bar Z_{kr} $ contribution.
As shown in \vall, the  
$Y_{jr}\bar\nabla Z^{rj} +
\bar Y^{rj}\nabla \bar Z_{jr}$ term contributes an anomaly
proportional to the dual coxeter number of the group, and
$Y_{jr}Z^{rk}\bar Z_{ks}\bar Y^{sj} -  
Z^{rj}Y_{js}\bar Y^{sk}\bar Z_{kr} $ contributes an anomaly
proportional to the level $k$ in the OPE of the Lorentz currents.
In the $AdS_5\times S^5$ case, the relevant group was 
$SO(4,1)\times SO(5)$ with dual coxeter number 3, which cancels the
level $k=-3$ in the OPE of the Lorentz currents constructed from pure
spinors \vall. In this case, the relevant group is $SU(2,2)\times SU(4)$
with dual coxeter number 4, which cancels the level $k=-4$
in the OPE of Lorentz currents constructed from unconstrained bosonic
spinors.

\subsec{Open string sector}

Just as d=3 Chern-Simons theory is described by the open string
sector of a topological A-model \csw, it will be shown that the open
string sector of the non-linear topological A-model 
of \actftwo\ describes N=4 d=4 super-Yang-Mills.
The open string boundary condition for the A-model of \actftwo\ will
be defined as 
\eqn\osb{\Tb_{jr} = \d_{jk}\e_{rs}\T^{sk}}
where $\e_{rs}$ is an antisymmetric tensor which breaks $SU(2,2)$ to
$SO(3,2)$ and $\d_{jk}$ is a symmetric tensor which breaks $SU(4)$ to
$SO(4)$. The boundary condition of \osb\ is similar to the open
string boundary condition for the Chern-Simons topological string which is
$\bar X_I = \d_{IJ} X^J$ for $I,J=1$ to 3. Note that the open string
boundary for the A-model is defined by 
\eqn\opendef{z=\bar z, \quad \k_+ = \bar \k_-, \quad \kb_+ = \k_-,}
so \osb\ implies that 
\eqn\osbtwo{\tb_{jr} = \d_{jk}\e_{rs}\t^{sk}, \quad 
\bar Z_{jr} = \d_{jk}\e_{rs}Z^{sk},\quad 
Y_{jr} = \d_{jk}\e_{rs}\bar Y^{sk}.} 
The boundary condition of \osb\ breaks half of the fermionic isometries
and reduces the $SU(2,2|4)$ supergroup of isometries
to the supergroup $OSp(4|4)$. This supergroup contains
$SO(3,2)\times SO(4)$ bosonic isometries and 16 fermionic isometries,
and is the N=4 supersymmetry algebra on $AdS_4$.

To show that the 
BRST cohomology of open string states in this model
describes N=4 d=4 super-Yang-Mills, it will be assumed that, as in
the topological A-model for Chern-Simons, the cohomology in the
closed string sector is trivial. 
This assumption is reasonable since N=(2,2)
worldsheet supersymmetric D-terms are BRST-trivial, and there are naively no
global obstructions to writing supersymmetric expressions involving
fermionic superfields as superspace D-terms.
However, since the A-model of \actftwo\ is constructed from fermionic
superfields in a non-conventional manner, there might be unexpected
subtleties in the model which invalidate this assumption.

With this assumption, the cohomology computation in the open
string sector is independent of $\L$ and $R$ in \actftwo, and
can be performed at
$\Lambda=0$ where only the constant modes of $\T^{rj}$ contribute. 
Furthermore, if the closed string sector has no cohomology, the
open string physical states should be independent of $SU(2,2|4)/OSp(4|4)$
rotations which modify the D-brane boundary conditions of \osb. So
although only $OSp(4|4)$ symmetry is manifest in the open topological
A-model, the physical spectrum should be invariant under the full
$SU(2,2|4)$ supergroup.

After imposing the open string boundary condition of \osb\ and 
restricting to constant worldsheet modes, the superspace action of
\actftwo\ reduces to
\eqn\redft{S= R^2 \int d\tau d^2 \k
~Tr [D_+\T (1+\T\T)^{-1} D_-\T (1+\T\T)^{-1}]}
where $\T_{jr} = \d_{jk}\e_{rs} \T^{sk}$ is an N=2 superfield whose
component expansion is
\eqn\compl{\T^{rj}=\t^{jr} + \k_+ Y^{rj} + \k_- Z^{rj} + \k_+\k_- f^{rj},}
and $D_\pm = {\p\over{\k^{\pm}}} + \k^\mp \dt.$
Alternatively, using the coset construction, the action of \cosetp\
reduces to
\eqn\cosetr{S = 
 R^2\int d\tau
 [\e_{rs}J^{rj}J^{sj} +(J-{\cal A})_s^r (J-{\cal A})_r^s
 -(J-{\cal A})_j^k (J-{\cal A})_k^j + 
Y_{jr} (\dt Z+ {\cal A}Z)^{rj} ] }
$$=
R^2\int d\tau
[\e_{rs}J^{rj}J^{sj} + 
Y_{jr}(\nabla Z)^{rj} 
+ (Y Z)_j^k(YZ )^j_k - (YZ)_r^s(YZ)^r_s],$$
where $J^A= (G^{-1} \dt G)^A$ are left-invariant currents taking
values in the Lie algebra of $OSp(4|4)$, 
$G(\t)$ 
takes values in the coset ${OSp(4|4)}\over{SO(3,2)\times SO(4)}$, 
$A= ([rs],[jk], jr)$ labels
the $OSp(4|4)$ generators, $r=1$ to 4 labels $Sp(4)$
indices which are raised and lowered using
the antisymmetric metric $\e^{rs}$, $j=1$ to 4 labels $SO(4)$
indices which are raised and lowered using $\d_{jk}$, 
${\cal A}^A$ is an $Sp(4)\times SO(4)$ worldline gauge field, and
$(\nabla Z)^{rj}=\dt Z^{rj} + J^r_s Z^{sj} + J^j_k Z^{rk}$.
The N=2 worldline supersymmetry generators for this action are
\eqn\wlsu{Q = Z^{rj} J_{jr}, \quad b = Y_{jr} J^{rj}.}

\newsec{Cohomology of Open Topological A-Model}

Before showing that the BRST cohomology of the worldline action of \cosetr\
describes N=4 d=4 super-Yang-Mills, it will be useful to review
the superspace description of on-shell super-Yang-Mills.

\subsec{On-shell super-Yang-Mills in superspace}

In ten flat dimensions, on-shell super-Yang-Mills is described
by a spinor superfield $A_\a(x,\t)$ where $\a=1$ to 16. This superfield
can be understood as a spinor connection which covariantizes the
superspace derivative $D_\a = {\p\over{\p\t^\a}} + \g_{\a\b}^c 
{\p\over{\p x^c}}$ to $\nabla_\a = D_\a - A_\a(x,\t)$. 
Since $\{D_\a, D_\b\}= \g^c_{\a\b}
{\p\over{\p x^c}}$, it is natural to impose that $A_\a$ is defined
such that \ref\siegelsuper{W. Siegel, {\it 
Superfields in higher-dimensional spacetime},
Phys. Lett. B80 (1979) 220.}
\eqn\bianchi{\{\nabla_\a, \nabla_\b\}= \g^c_{\a\b}\nabla_c }
where
$\nabla_c = 
{\p\over{\p x^c}} - A_c(x,\t)$ 
and $A_c(x,\t)$ is a vector connection whose $\t=0$ component is the
usual gauge field. 

These spinor and vector superspace connections are defined up to
the gauge transformation
\eqn\gaugetran{\d A_\a = \nabla_\a \O, \quad \d A_c = \nabla_c \O}
where $\O$ is a scalar superfield, and the Bianchi identity of
\bianchi\ implies that 
\eqn\eqone{D_\a A_\b + D_\b A_\a - \{A_\a,A_\b\} = \g^c_{\a\b} A_c.}
Equation \eqone\ implies that $A_c$ is determined from $A_\a$ and
that $A_\a$ must satisfy the constraint
\eqn\eqtwo{(\g^{abcde})^{\a\b}(D_\a A_\b - \half\{A_\a,A_\b\})=0}
for
any five-form direction $abcde$ \ref\howepure{P. Howe,
{\it Pure spinors lines in superspace and ten-dimensional
supersymmetric theories}, Phys. Lett. B258 (1991) 141.}.

The constraint of \eqtwo\ together with the gauge invariance of \gaugetran\
implies that $A_\a(x,\t)$ can be gauged to the form
\eqn\formaa{A_\a(x,\t) = a_c(x) (\g^c \t)_\a + \xi^\b(x) (\g^c\t)_\b
(\g_c\t)_\a + ...}
where $a_c(x)$ and $\xi^\a(x)$ are the on-shell gluon and gluino, and
$...$ involves spacetime derivatives of $a_c(x)$ and $\xi^\a(x)$.

To describe N=4 d=4 super-Yang-Mills, one simply decomposes the
d=10 vectors and spinors into d=4 vectors, scalars and spinors in 
the usual manner as
\eqn\decom{\t^\a \to (\t^{\mu j}, \tb^{\md}_j), \quad
A_\a \to (A_{\mu j}, \bar A_\md^j), \quad A_c \to (A_m, A_{[jk]})}
where $m=0$ to 3, $\mu,\md=1$ to 2, $j=1$ to 4, and $[jk]=1$ to 6.
The corresponding covariant spinor and vector derivatives
satisfy the
Bianchi identities
\eqn\bianchitwo{\{\nabla_{\mu j},\bar \nabla_\md^k\} = \d_j^k \s^m_{\mu\md}
\nabla_m, \quad
\{\nabla_{\mu j}, \nabla_{\nu k}\} = \e_{\mu\nu} A_{[jk]}, \quad
\{\bar\nabla_\md^j, \bar\nabla_\nd^k\} =\half \e_{\md\nd}\e^{hijk} A_{[hi]},}
where $\s^m_{\mu\md}$ are the d=4 Pauli matrices.
So the N=4 d=4 spinor connections satisfy the equations
\eqn\eqfour{D_{\mu j} \bar A_{\nd}^k + \bar 
D_\nd^k A_{\mu j} - \{A_{\mu j},\bar A_\nd^k\} 
= \d_j^k \s^m_{\mu \nd} A_m,}
$$
D_{(\mu j} A_{\nu k)} - 
\{A_{\mu j},A_{\nu k}\}   
 = \e_{\mu\nu} A_{[jk]}, \quad
\bar D^{(\md j}\bar  A^{\nd k)} - 
\{\bar A^{\md j},\bar A^{\nd k}\}   
 = \half\e^{\md\nd} \e^{hijk} A_{[hi]},$$
and the gauge transformations
\eqn\gaugefour{\d A_{\mu j}= \nabla_{\mu j}\O, \quad
\d \bar A_{\md}^j= \bar \nabla_{\md}^j\O, \quad \d A_m = \nabla_m\O.}

Since N=4 d=4 super-Yang-Mills is superconformally invariant, the
Bianchi identities of \bianchitwo\ are valid both in flat d=4
Minkowski space and in $AdS_4$ space. The only difference is that
in a flat background, the superspace derivatives are 
\eqn\flatd{D_{\mu j} = {\p\over{\p\t^{\mu j}}} + \tb^\md_j \s^m_{\mu\md}
{\p\over{\p x^m}}, \quad
\bar D_{\md}^j = {\p\over{\p\tb^{\md}_j}} + \t^{\mu j} \s^m_{\mu\md}
{\p\over{\p x^m}}, \quad
D_m = 
{\p\over{\p x^m}}, }
whereas in an $AdS_4$ background, 
\eqn\adsd{D_A = E_A^M 
{\p\over{\p Y^M}} + w_{A}^{[mn]} M_{[mn]} + w_{A}^{[jk]} M_{[jk]}}
where $E_A^M$ is the $AdS_4$ super-vierbein, $Y^M = (y^m,\xi^{\mu j},
\bar\xi^{\md}_j)$ are the $AdS_4$ superspace coordinates, 
$w_A$ is the $AdS_4$ super-connection, and $M_{[mn]}$ and
$M_{[jk]}$ are the $SO(3,1)$ and $SO(4)$ generators.
As will be shown in subsection 5.3, the $AdS_4$ super-vierbein
and super-connection can be naturally constructed from 
a supercoset ${OSp(4|4)}\over{SO(3,1)\times SO(4)}$ in the same manner as
the $AdS_5\times S^5$ super-vierbein and super-connection are
constructed from the ${PSU(2,2|4)}\over{SO(4,1)\times SO(5)}$ supercoset.

\subsec{First-quantized description of N=4 d=4 super-Yang-Mills}

Just as d=3 Chern-Simons can be obtained by quantizing
the worldline action $\int d\tau (\half {{\p x^I}\over{\p \tau}} 
{{\p x_I}\over {\p \tau}} + \bar\psi_I
\dt\psi^I)$ with the BRST operator $Q= \psi^I \dt x_I$ where $I=1$ to 3,
d=10 super-Yang-Mills can be obtained by quantizing the worldline
action
$\int d\tau (\half {{\p x^c}\over{\p \tau}}{{\p x_c}\over{\p\tau}}
+ p_\a \dt \t^\a + w_\a \dt \l^\a)$
with the BRST operator $Q= \l^\a d_\a$ where
$d_\a = p_\a + (\g_c\t)_\a \dt x^c$ and $\l^\a$ is a pure spinor
satisfying $\l\g^c\l=0$ for $c=0$ to 9 \spar\howepure.

At ghost-number one, the states in the cohomology of $Q= \l^\a d_\a$
are described by $V= \l^\a A_\a(x,\t)$
where $A_\a(x,\t)$ is a spinor superfield. $QV=0$ implies that
$\l^\a\l^\b D_\b A_\a=0$ where $D_\a = {\p\over{\p\t^\a}}
+ (\g^c\t) {\p\over{\p x^c}}$, and since $\l\g^c\l=0$, 
$\l^\a\l^\b D_\b A_\a=0$ implies that 
$D_\a A_\b + D_\b A_\a = \g^c_{\a\b} A_c$ for some $A_c$.
Also, $\d V= Q\O$ implies that $\d A_\a= D_\a \O$. By comparing
with \eqone\ and \gaugetran, one sees that $A_\a(x,\t)$ describes
the linearized on-shell d=10 super-Yang-Mills fields.

The structure of $V=\l^\a A_\a(x,\t)$ in d=10 super-Yang-Mills
using the BRST operator $Q= \l^\a d_\a$ closely
resembles the structure of $V=\psi^I A_I(x)$ in Chern-Simons theory using
the BRST operator $Q= \psi^I \dt x_I$. In Chern-Simons theory,
$QV=0$ implies that $\p_I A_J - \p_J A_I =0$ and $\d V = Q\O$
implies that $\d A_I = \p_I \O$. Furthermore, as in Chern-Simons theory,
the super-Yang-Mills ghost is described by the BRST cohomology at ghost-number
zero, 
the super-Yang-Mills fields are described by the BRST 
cohomology at ghost-number one, 
the super-Yang-Mills antifields are described by the BRST 
cohomology at ghost-number two, and the super-Yang-Mills antighost
is described by the BRST cohomology at ghost-number three \spar.
This structure can be seen from the Batalin-Vilkovisky action
for d=10 super-Yang-Mills which can be written in the Chern-Simons-like
form $S= \langle V Q V + {2\over 3}V^3 \rangle$ using the normalization
convention that 
$\langle (\l\g^a\t)(\l\g^b\t)(\l\g^c\t)(\t\g_{abc}\t)\rangle =1$.

This construction for d=10 super-Yang-Mills
is easily generalized to N=4 d=4 super-Yang-Mills
by eliminating six of the ten $x$'s and decomposing the d=10 spinors
into N=4 d=4 spinors as
\eqn\decomtwo{
\t^\a \to (\t^{\mu j}, \tb^{\md}_j), \quad
p_\a \to (p_{\mu j}, \bar p_\md^j), \quad 
\l^\a \to (\l^{\mu j}, \bar\l^{\md}_j), \quad
w_\a \to (w_{\mu j}, \bar w_\md^j),}
where $\mu,\md=1$ to 2 and $j=1$ to 4.
The pure spinor condition $\l\g^c\l=0$ implies that 
$\l^{\mu j}$ and $\bar\l^{\md}_j$ satisfy the constraints
\eqn\fourcon{\l^{\mu j} \bar\l^\md_j =0,}
\eqn\fourcont{\e_{\mu\nu}\l^{\mu j}\l^{\nu k}
= \half\e_{\md\nd} \e^{hijk} \bar\l^\md_h \bar\l^\nd_i.}
Although \fourcon\ and \fourcont\ contain ten constraints, only five of these
constraints are independent. This is easy to verify since
$\l^{ \mu j} \bar\l^\md_j =0$ implies that 
$\bar\l^{\dot\rho}_j (\e_{\mu\nu} \l^{\mu j}\l^{\nu k}) =0$, which implies that
\eqn\impl{\e_{\mu\nu} \l^{\mu j}\l^{\nu k} = \half e^{2\phi} \e^{hijk}
\e_{\md\nd}\bar\l_h^\md\bar\l_i^\nd}
for some $\phi$. So if the four constraints in \fourcon\ are satisfied,
any one of the constraints in \fourcont\ imply that $\phi=0$, which implies
that the remaining five constraints in \fourcont\ are satisfied.

Since the four constraints of \fourcon\ are almost strong enough to
define an N=4 d=4 pure spinor, it
will be convenient to define a ``semi-pure'' spinor $(\lt^{\mu j},
\lbt^\md_j)$ which is only required to satisfy the four constraints of
\fourcon\ that 
\eqn\tilcon{\lt^{\mu j} \lbt^{\md}_j =0.}
A semi-pure spinor has 12 independent components and is related
to a pure spinor $(\l^{\mu j}, \bar\l^\md_j)$ 
by a $U(1)$ ``$R$-transformation'' as
\eqn\rrel{\lt^{\mu j} = e^{\phi\over 2}\l^{\mu j}, \quad
\lbt^{\md}_j = e^{-{\phi\over 2}}\bar\l^{\md}_j}
where $\phi$ is determined from 
\eqn\defphi{ e^{2\phi} = 
{{\e_{\mu\nu} \lt^{\mu j}\lt^{\nu k}}\over {\half\e^{hijk}
\e_{\md\nd}\lbt_h^\md\lbt_i^\nd}}.}

In flat d=4 Minkowski space, the worldline action for N=4 d=4
super-Yang-Mills will be defined as 
\eqn\flatwl{
S=\int d\tau (\half {{\p x^m}\over{\p \tau}}{{\p x_m}\over{\p \tau}}
+ p_{\mu j} \dt \t^{\mu j}
+ \bar p_{\md}^j \dt \tb^{\md }_j
+ \wt_{\mu j} \dt \lt^{\mu j}
+ \wbt_{\md}^j \dt \lbt^{\md }_j ]}
with the BRST operator
\eqn\flatbrst{Q = \lt^{\mu j} d_{\mu j} + \lbt^\md_j \bar d_\md^j}
where $d_{\mu j} = p_{\mu j} +\s^m_{\mu\md}\tb^\md_j {{\p x_m}\over{\p \tau}}$,
$\bar d_{\md }^j =\bar  p_{\md}^j +\s^m_{\mu\md}\t^{\mu j} 
{{\p x_m}\over{\p \tau}}$,
and $\lt^{\mu j}$ and
$\lbt^{\md}_j$ are semi-pure spinors satisfying \tilcon.
Note that $Q^2 =0$ since $\{d_{\mu j},\bar d_\md^k\}= \d_j^k
\s^m_{\mu\md}{{\p x_m}\over{\p\tau}}$, and that
$\wt_{\mu j}$ and $\wbt_\md^j$ can only appear in combinations which are
invariant under the gauge transformations
\eqn\gaugeww{\d \wt_{\mu j}= \xi_m \s^m_{\mu\md} \lbt^\md_j, \quad
\d \wbt_{\md}^j= \xi_m \s^m_{\mu\md} \lt^{\mu j}.}

The action and BRST operator of \flatwl\ and \flatbrst\ are
invariant under the $U(1)$ $R$-transformation
\eqn\rtr{\t^{\mu j}\to c\t^{\mu j}, \quad
\tb^{\md}_{ j}\to c^{-1}\tb^{\md}_{ j}, \quad
p_{\mu j} \to c^{-1} p_{\mu j}, \quad
\bar p_{\md}^{ j} \to  c\bar p_{\md}^{ j}, }
$$\lt^{\mu j}\to c\lt^{\mu j}, \quad
 \lbt^{\md}_{ j}\to   c^{-1}\lbt^{\md}_{ j}, \quad
\wt_{\mu j} \to c^{-1} \wt_{\mu j}, \quad
\wbt_{\md}^{ j} \to  c \wbt_{\md}^{ j}, $$
however, N=4 d=4 super-Yang-Mills
does not contain such a $U(1)$ symmetry.
Since the variable $\phi$ of \defphi\ transforms under \rtr\ as 
\eqn\phitr{\phi \to \phi + \half \log ~c,}
$\phi$ can be interpreted as a ``compensator'' for $U(1)$ $R$-transformations
which cancels the $U(1)$ $R$-transformation of $\t^{\mu j}$ and $\tb^\md_j$.
Physical states will therefore be defined as states of
$+1$ ghost-number in the BRST cohomology which are invariant
under the $R$-transformation of \rtr. 

At ghost-number one, $R$-invariant
states are described by
\eqn\statesa{V = e^{-{\phi\over 2}}\lt^{\mu j} A_{\mu j}(x,
\t e^{-{\phi\over 2}}, \tb e^{\phi\over 2}) + e^{\phi\over 2}
\lbt^{\md}_j \bar A_\md^j(x,\t e^{-{\phi\over 2}},\tb e^{\phi\over 2})}
where $\phi$ is defined in \defphi\ and cancels the $R$-transformation
of $\lt$ and $\t$. In other words, 
\eqn\statec{V = \lt^{\mu j} A'_{\mu j}(x,\t',\tb') +
 \lbt^{\md }_j \Abt_{\md }^j (x,\t',\tb')}
where  
$A'_{\mu j}(x,\t',\tb')=
e^{-{\phi\over 2}}A_{\mu j}(x, \t e^{-{\phi\over 2}}, \tb e^{\phi\over 2})$
and
$\Abt_{\md }^j (x,\t',\tb') =
e^{\phi\over 2}
\bar A_\md^j(x,\t e^{-{\phi\over 2}},\tb e^{\phi\over 2})$ are the
$R$-transformed versions of 
$A_{\mu j}(x,\t,\tb)$ and
$\bar A_{\md }^j (x,\t,\tb)$ using the $R$-parameter $c=e^{-{\phi\over 2}}$
in \rtr.
The equation $QV=0$ implies that 
\eqn\expone{e^{-\phi} \lt^{\mu j}\lt^{\nu k}
D_{\mu j} A_{\nu k} + e^{\phi} \lbt^\md_j\lbt^\nd_k \bar D_\md^k \bar A_\nd^k
+ \lt^{\mu j} \lbt_k^\nd (D_{\mu j}\bar A_\nd^k +\bar D_{\nd}^k A_{\mu j}) =0,}
which implies 
using the pure spinor constraints of \fourcon\ - \defphi\ that
\eqn\imptwo{D_{\mu j} \bar A_\nd^k + \bar D_\nd^k A_{\mu j} = 
\d_j^k \s_{\mu\nd}^m A_m,
\quad
D_{(\mu j} A_{\nu k)} = \e_{\mu\nu} A_{[jk]},\quad
\bar D^{(\md j} \bar A^{\nd k)} = \half \e^{\md\nd} \e^{hijk} A_{[hi]},}
for some superfields $A_m(x,\t,\tb) $ and 
$A_{[jk]}(x,\t,\tb)$. Furthermore, the gauge transformation $\d V=Q
\O (x, e^{-{\phi\over 2}}\t, e^{\phi\over 2}\tb)$ implies that 
\eqn\furtg{\d A_{\mu j} = D_{\mu j} \O, \quad
\d \bar A_{\md}^j = \bar D_{\md}^j \O, \quad
\d A_{m} = \p_m \O.}

So when $V$ of \statesa\ is in the BRST cohomology,
$A_{\mu j}$ and $\bar A_\md^j$ satisfy
the linearized N=4 d=4 super-Yang-Mills 
equations and gauge invariances of \eqfour\ and \gaugefour\
in flat Minkowski space.

\subsec{N=4 d=4 super-Yang-Mills in $AdS_4$}

To generalize this construction to N=4 d=4 super-Yang-Mills in
an $AdS_4$ background, one needs to modify the worldline action
and BRST operator of \flatwl\ and \flatbrst\ to be $OSp(4|4)$
invariant. This can be done using a coset construction based on
${OSp(4|4)}\over {SO(3,1)\times SO(4)}$ which contains four
bosonic generators and sixteen fermionic generators. As in the
$AdS_5\times S^5$ construction, it is convenient to define 
left-invariant currents $J^A = (g^{-1} \dt g)^A$ where
$g(x,\t)$ takes values in the 
${OSp(4|4)}\over {SO(3,1)\times SO(4)}$ coset,
$A= (m,[mn],[jk], r j)$ label the $OSp(4,4)$
generators, $m=0$ to 3 label the ``translation'' generators, 
$[mn]$ and $[jk]$ label the $SO(3,1)$ and $SO(4)$ generators,
and $r j$ label the ``supersymmetry'' generators for 
$r=1$ to 4 and $j=1$ to 4. Note that the two-component $\mu$ index
corresponds to $r=1,2$, the two-component $\md$ index corresponds
to $r=3,4$, and the antisymmetric $\e_{rs}$ tensor has non-zero
components $\e_{12}=-\e_{21}=\e_{34}=-\e_{43}=1$.

The $OSp(4|4)$-invariant worldline action is
\eqn\wlo{S = R^2 \int d\tau [{1\over 4}J^m J_ m + \e_{rs} J^{r j} J^{s j} 
+ \wt_{r j} (\dt\lt + {\cal A}\lt)^{rj}}
$$+ (J^{[mn]}-{\cal A}^{[mn]})
(J_{[mn]}-{\cal A}_{[mn]})
- (J^{[jk]}-{\cal A}^{[jk]})
(J_{[jk]}-{\cal A}_{[jk]})]$$
$$= 
R^2 \int d\tau [{1\over 4}J^m J_ m + \e_{rs} J^{r j} J^{s j} 
+ \wt_{r j}(\nabla \lt)^{r j} + (\wt\lt)_j^k(\wt\lt)_k^j
- (\wt\s^{mn}\lt) (\wt\s_{mn}\lt)],$$
where $(\wt\lt)_j^k = \wt_{rj}\lt^{rk}$,
 $(\wt\s^{mn}\lt) = (\s^{mn})_s^r\wt_{rj} \lt^{sj} $ and
$(\nabla \lt)^{rj} = \dt\lt^{rj} +\half J_{[mn]}(\s^{[mn]})_s^r \lt^{sj} +
J^j_k \lt^{rk}.$
This action
is invariant under local $SO(3,1)\times SO(4)$ transformations
where $\lt$ and $\wt$ transform covariantly, and is also
invariant under the BRST transformations
\eqn\brtr{\d g = g (\e \lt^{r j} T_{r j}), \quad \d \wt_{r j} = \e J_{r j},} 
generated by the BRST operator
$Q = \lt^{r j} J_{r j}$
where $T_{rj}$ are the fermionic generators of $OSp(4|4)$.

Defining the ghost-number one vertex operator as
\eqn\defva{V = \lt^{rj} \At_{rj} = \lt^{\mu j} \At_{\mu j} + \lbt^{\md}_j
\Abt_\md^j,}
the BRST-transformation of \brtr\ implies that
\eqn\qva{ QV = \lt^{\mu j}\lt^{\nu k} 
\nabla_{\mu j} \At_{\nu k} + \lbt^\md_j\lbt^\nd_k \bar \nabla_\md^k \Abt_\nd^k
+ \lt^{\mu j} \lbt_k^\nd 
(\nabla_{\mu j}\Abt_\nd^k +\bar \nabla_{\nd}^k A'_{\mu j}),}
where $\nabla_{\mu j}$ and $\bar\nabla_\md^j$ are the covariant superspace
derivatives in an $AdS_4$ background.
So $QV=0$ implies that 
\eqn\impntwo{\nabla_{\mu j} \Abt_\nd^k + \bar \nabla_\nd^k \At_{\mu j} = 
\d_j^k \s_{\mu\nd}^m A_m,
\quad
e^\phi \nabla_{(\mu j} A'_{\nu k)} = \e_{\mu\nu} A_{[jk]},\quad
e^{-\phi}
\bar \nabla^{(\md j} \Abt^{\nd k)} = \half \e^{\md\nd} \e^{hijk} A_{[hi]},}
for some superfields $A_m$ and 
$A_{[jk]}$.

Although the equations of
\impntwo\ are difficult to solve when written in terms
of $AdS_4$ superspace variables, they can be simplified by performing
a superconformal transformation from N=4 $AdS_4$ superspace into
N=4 d=4 Minkowski superspace. A point $(y^m,\xi^{\mu j}, \bar\xi^\md_j)$
in $AdS_4$ superspace can be represented as 
\eqn\repads{g_{AdS_4}(y,\xi,\bar\xi) 
= e^{y^m (P_m + K_m) + \xi^{\mu j}
(Q_{\mu j} + S_\mu^k \d_{jk}) + \bar\xi^\md_j
(\bar Q_{\md}^j + \bar S_{\md k} \d^{jk}) }}
where $g(y,\xi,\bar\xi)$ is an element of $PSU(2,2|4)$ whose bosonic generators
for translations, conformal boosts, rotations, dilatations and $SU(4)$
$R$-transformations are denoted respectively by $[P_m,K_m, M_{[mn]}, D, R_j^k]$,
and whose fermionic generators for supersymmetry and superconformal
transformations are denoted respectively by $[Q_{ \mu j}, \bar Q_\md^j,
S_\mu^j, \bar S_{\md j}]$.
Under an N=4 superconformal transformation parameterized by the $PSU(2,2|4)$
element $\Omega$, 
\eqn\traone{g_{AdS_4}(y,\xi,\bar\xi) \to g'_{AdS_4}
(y',\xi',\bar\xi')= \Omega ~ g_{AdS_4}(y,\xi,\bar\xi)~ h (y,\xi,\bar\xi)}
where
\eqn\defhh{h = e^{c^m K_m + w^{mn} M_{[mn]} + a^j_k R_j^k + b D
+ \chi^\mu_j S^j_\mu + \bar \chi^{\md j} \bar S_{\md j}}}
and the parameters $[c^m, w^{mn}, a^j_k, b, \chi^\mu_j,\bar\chi^\md_j]$
in \defhh\ are chosen such that 
\eqn\chosen{g'_{AdS_4} = 
e^{ y'^m (P_m + K_m) + \xi'^{\mu j}
(Q_{\mu j} + S_\mu^k \d_{jk}) + \bar\xi'^\md_j
(\bar Q_{\md}^j + \bar S_{\md k} \d^{jk}) }}
for some $(y'^m(y,\xi,\bar\xi), 
\xi'^{\mu j}(y,\xi,\bar\xi), \bar\xi'^\md_j(y,\xi,\bar\xi))$.

Similarly, a point $(x^m, \t^{\mu j}, \tb^\md_j)$ in N=4 d=4
Minkowski superspace can be represented as
\eqn\repadt{g_{Mink}(x,\t,\tb) = e^{x^m P_m  + \t^{\mu j}
Q_{\mu j} + \tb^\md_j
\bar Q_{\md}^j}}
where under an N=4 
superconformal transformation parameterized by 
$\Omega$, 
\eqn\tratwo{g_{Mink}(x,\t,\tb) \to g'_{Mink}(x', \t', \tb') = \Omega ~
g_{Mink} (x,\t,\tb)~h (x,\t,\tb)}
and
the parameters $[c^m, w^{mn}, a^j_k, b, \chi^\mu_j,\bar\chi^\md_j]$
in $h$ of \defhh\ are now chosen such that $g'_{Mink} = 
e^{x'^m P_m  + \t'^{\mu j}
Q_{\mu j} + \tb'^\md_j
\bar Q_{\md}^j }$
for some $(x'^m(x,\t,\tb), \t'^{\mu j}(x,\t,\tb), \tb'^\md_j(x,\t,\tb))$.

To superconformally map N=4 $AdS_4$ superspace into N=4 d=4 Minkowski 
superspace,
define
\eqn\confm{g_{Mink}(x,\t,\tb) = g_{AdS_4}(y,\xi,\bar\xi) ~h(y,\xi,\bar\xi)}
where the parameters $[c^m, w^{mn}, a^j_k, b, \chi^\mu_j,\bar\chi^\md_j]$
in $h$ of \defhh\ are chosen such that $g_{Mink} = 
e^{x^m P_m + \t^{\mu j}
Q_{\mu j} + \tb^\md_j
\bar Q_{\md}^j }$
for some functions $(x^m(y,\xi,\bar\xi),
\t^{\mu j}(y,\xi,\bar\xi), \tb^\md_j( y,\xi,\bar\xi))$.
After writing the $AdS_4$ superspace variables $(y^m, \xi^{\mu j}, 
\bar\xi^\md_j)$ in terms of the Minkowski superspace variables
$(x^m, \t^{\mu j}, 
\tb^\md_j)$ using this superconformal map, 
the superfield equations of \impntwo\
simplify to
\eqn\imptr{D_{\mu j} \Abt_\nd^k + \bar D_\nd^k \At_{\mu j} = 
\d_j^k \s_{\mu\nd}^m A_m,
\quad
e^{\phi} D_{(\mu j} A'_{\nu k)} = \e_{\mu\nu} A_{[jk]},\quad
e^{-\phi}\bar D^{(\md j} \Abt^{\nd k)} =  \half\e^{\md\nd} \e^{hijk} A_{[hi]},}
where $D_{\mu j}$ and $\bar D_\md^j$ are the flat superspace derivatives.
So if one defines 
$A'_{\mu j}(x,\t',\tb')=
e^{-{\phi\over 2}}A_{\mu j}(x, \t e^{-{\phi\over 2}}, \tb e^{\phi\over 2})$
and
$\Abt_{\md }^j (x,\t',\tb') =
e^{\phi\over 2}
\bar A_\md^j(x,\t e^{-{\phi\over 2}},\tb e^{\phi\over 2})$ as in
\statec, one finds that 
\eqn\impts{D_{\mu j} \bar A_\nd^k + \bar D_\nd^k A_{\mu j} = 
\d_j^k \s_{\mu\nd}^m A_m,
\quad
D_{(\mu j} A_{\nu k)} = \e_{\mu\nu} A_{[jk]},\quad
\bar D^{(\md j} \bar A^{\nd k)} =  \half\e^{\md\nd} \e^{hijk} A_{[hi]},}
which are the same equations as 
\imptwo. So the $OSp(4|4)$-invariant worldline action of 
\wlo\ also describes N=4 d=4 super-Yang-Mills.

\subsec{Equivalence with open topological A-model}

It will now be shown that the worldline action of \wlo, which
is based on the 
${OSp(4|4)}\over {SO(3,1)\times SO(4)}$ coset together with semi-pure
spinors, is related by a field redefinition to the worldline action
of \cosetr, which is based on the 
${OSp(4|4)}\over {SO(3,2)\times SO(4)}$ coset together with unconstrained
spinors. This field redefinition combines the four $x$'s of the
${OSp(4|4)}\over {SO(3,1)\times SO(4)}$ coset with the 12 components
of the semi-pure spinors to form an unconstrained 16-component
spinor which transforms covariantly like a twistor variable under
$SO(3,2)$ transformations. The construction of this $AdS_4$
twistor variable is very similar to the construction of the
$AdS_5\times S^5$ twistor variable of subsection 3.2 in which
the ten $x$'s of the ${PSU(2,2|4)}\over{SO(4,1)\times SO(5)}$
coset were combined with the 22 components of the pure spinors to
form two unconstrained 16-component spinors.

To construct the field redefinition, first decompose the
${OSp(4|4)}\over{SO(3,1)\times SO(4)}$
coset as
\eqn\decomposek{g(x,\t) = e^{\t^{rj} T_{rj}} e^{x^m T_m}
 \equiv  G(\t) H(x)}
where $G(\t) = e^{\t^{rj} T_{rj}}$ takes values in
${OSp(4|4)}\over{Sp(4)\times SO(4)}$, $H(x) = e^{x^m T_m}$
takes values in ${Sp(4)}\over {SO(3,1)}$,
and $T_{rj}$ and $T_m$ are the ``supersymmetry'' and ``translation''
generators of ${OSp(4|4)}\over {SO(3,1)\times SO(4)}$.

Now define the twistor-like variable as
\eqn\twist{Z^{rj } = H^r_s\lt^{s j}}
which combines the four $x$'s in $H^r_s$ with the 12 components
of the semi-pure spinor
$\lt$. Similarly, define the conjugate twistor-like variable as
\eqn\twisty{Y_{jr } = (H^{-1})^s_r\wt_{j s}.}
Using 
\eqn\cose{J = (g^{-1}\dt g) = (H^{-1}\dt H) + H^{-1} (G^{-1}\dt G) H,}
one finds that 
\eqn\relf{Y_{jr}\dt Z^{rj} = \wt_{rj}\dt \lt^{rj}
+ (H^{-1} \dt H)_r^s (\wt\lt)^r_s}
$$ = 
\wt_{rj}\dt \lt^{rj}
+ J_r^s (\wt\lt)^r_s
- (G^{-1}\dt G)_r^s (Y Z)^r_s $$
$$ = \wt_{rj} (\nabla \lt)^{rj} + J^m (\wt\s_m \lt)
- (G^{-1}\dt G)_r^s (Y Z)^r_s
- (G^{-1}\dt G)_j^k (Y Z)^j_k ,$$
where $(\wt\lt)^r_s= \wt_{js}\lt^{rj}$,
$(\wt\lt)^j_k= (YZ)^j_k = Y_{kr}Z^{rj}$, 
$(\wt\s^m\lt) = (\s^m)^r_s \wt_{ rj } \lt^{ sj}$, and
$(\nabla \lt)^{rj}= \dt\lt^{rj} +\half J^{mn}(\s_{mn}\lt)^{rj}
+ J^j_k \lt^{rk}$. 
Furthermore,  
\eqn\sqw{
 (\wt\s^{mn}\lt) (\wt\s_{mn}\lt) 
=
(\wt\lt)_r^s(\wt\lt)^r_s
-(\wt\s^m\lt)(\wt\s_m\lt)}
$$ = 
(YZ)_r^s(YZ)^r_s
-(\wt\s^m\lt)(\wt\s_m\lt).$$

Plugging \relf\ and \sqw\ into the action of \wlo, and introducing
an auxiliary variable $P_m$ to write the $J_m J^m$ kinetic term in
first-order form, one finds that the action of \wlo\ can be written as
\eqn\wltwo{S = \int d\tau [P_m J^m - P_m P^m + \e_{rs} J^{rj} J^{sj}
+ Y_{jr}(\nabla Z)^{rj} }
$$+ (Y Z)_j^k(YZ )^j_k - (YZ)_r^s(YZ)^r_s - J^m (\wt\s_m \lt) +
(\wt\s^m\lt)(\wt\s_m\lt)]$$
$$= 
 \int d\tau [P'_m (J^m -2 \wt\s^m \lt) - P'_m P'^m + \e_{rs} J^{rj} J^{sj}
+ Y_{jr}(\nabla Z)^{rj} 
+ (Y Z)_j^k(YZ )^j_k - (YZ)_r^s(YZ)^r_s],$$
$${\rm where ~~~}(\nabla Z)^{rj}
=\dt Z^{rj} +  (G^{-1}\dt G)_s^r Z^{sj}
+ (G^{-1}\dt G)_k^j Z^{rk} {\rm ~~~ and}$$
\eqn\pdef{P'_m = P_m - (\wt\s_m\lt).}

Under the gauge transformation $\d\wt_{rj} = \xi^m (\s_m)_r^s \lt_{sj}$
of \gaugeww, 
\pdef\ implies that 
\eqn\chap{\d P'_m = \xi^n (\s_{mn})_r^s \lt^{rj}\lt_{sj}.}
For generic values of $\lt^{rj}$, $\det (\d P'/\d\xi)$ is non-zero, so
one can consistently gauge $P'_m =0$. Moreover,
it is expected that the Fadeev-Popov
factor from this gauge-fixing of $P'_m$ 
is cancelled by the measure factor which
converts the four $x$'s and 12 constrained $\lt$'s into the 16 unconstrained
$Z^{rj}$'s.

In the gauge $P'_m=0$, the action of \wltwo\ reduces to 
\eqn\wltho{S =
\int d\tau [\e_{rs} J^{rj} J^{sj}
+ Y_{rj}(\nabla Z)^{rj} 
+ (Y Z)_j^k(YZ )^j_k - (YZ)_r^s(YZ)^r_s],}
where \cose\ implies that
$\e_{rs} J^{rj} J^{sj} = \e_{rs} (G^{-1} \dt G)^{rj}
(G^{-1} \dt G)^{sj}$. 
Since $G$ parameterizes the coset ${OSp(4|4)}\over{SO(3,2)\times SO(4)}$,
the worldline action of \wltho\ is equivalent to the worldline action of
\cosetr\ coming from the open topological A-model. And since the
BRST cohomology of \wlo\ describes d=4 N=4 super-Yang-Mills, this
equivalence implies that the physical states in the open sector
of the topological A-model are d=4 N=4 super-Yang-Mills states.

\newsec{Conclusions}

In this paper, a new limit of the $AdS_5\times S^5$ sigma model
was considered in which the vector components of the $PSU(2,2|4)$
metric $g_{ab}\to \infty$ and the superspace torsion $T_{\a\b}{}^a\to 0$,
while the spinor 
components of the $PSU(2,2|4)$
metric $g_{\a\bh}$ and the superspace torsion $T_{\a a}{}^\bh$ are held
fixed. This is the opposite procedure from the flat space limit, and
if $(T_{\a\b}^b \eta_{ab})/
(T_{\a a}^\bh \eta_{\b\bh})$
is interpreted as the $AdS_5\times S^5$ radius, it corresponds
to taking this radius to zero. 

In this limit, the $PSU(2,2|4)$ algebra
deforms into an $SU(2,2)\times SU(4)$ bosonic algebra with 32
abelian fermionic isometries, and the $AdS_5\times S^5$ sigma
model reduces to a linear topological A-model constructed from
fermionic N=2 superfields. The bosonic components of these 
fermionic superfields involve twistor-like combinations of the
$x$'s and pure spinor ghosts, and the linear topological A-model
can be interpreted as the limit of a $PSU(2,2|4)$-invariant
non-linear topological A-model whose open
string sector describes
N=4 d=4 super-Yang-Mills. 

These results have many parallels with the open-closed duality
found by Gopakumar and Vafa which relates Chern-Simons theory
and the resolved conifold \gopv. In this open-closed duality, 
Chern-Simons theory is described by the open sector of a topological
A-model \csw, which is interpreted as a Coulomb branch of the closed string
theory for the resolved conifold. As pointed out in \gopv\ and \oov,
the Chern-Simons/conifold duality shares many features with the
Yang-Mills/$AdS_5\times S^5$ duality, suggesting that the 
Ooguri-Vafa worldsheet proof of Chern-Simons/conifold duality \oov\
might have a generalization to a worldsheet proof of the Maldacena
conjecture.

However, before attempting a proof of Maldacena's conjecture using
the results of this paper, one would need to understand better
both the properties of the $T_{\a\b}{}^a \to 0$ limit of the
$AdS_5\times S^5$ sigma model, and the properties of the
open topological A-model for N=4 d=4 super-Yang-Mills. 

For example,
it is not clear that the $T_{\a\b}{}^a\to 0$ limit of the sigma
model can be interpreted as the small $AdS_5\times S^5$ radius limit,
and that a separate Coulomb branch is developed in this limit.
Furthermore, although it was shown that the physical states of
the open topological A-model describes N=4 d=4 super-Yang-Mills,
it was not shown how to compute perturbative
super-Yang-Mills scattering amplitudes
using this A-model. Hopefully, the d=10 pure spinor formalism
will provide some useful clues for computing these amplitudes. For example,
if the d=10 pure spinor measure factor
$\langle (\l\g^a\t)(\l\g^b\t)(\l\g^c\t)(\t\g_{abc}\t)\rangle =1$
is dimensionally reduced to four dimensions,
the field theory action for the open A-model  
\eqn\sfta{S = \langle VQV + {2\over 3}V V V\rangle}
appears to correctly reproduce the N=4 d=4 super-Yang-Mills action \spar\wchw.
So using the interaction vertex from \sfta, it should be possible
to at least compute 3-point super-Yang-Mills tree amplitudes with
the open topological A-model. A much bigger challenge would be to
compute 4-point tree amplitudes using the A-model, and perhaps
the twistor-string methods of \wittwis\ref\twis{R. Roiban,
M. Spradlin and A. Volovich, {\it A googly amplitude from the B-model
in twistor space}, JHEP 0404 (2004) 012.}
\ref\altern{N. Berkovits, {\it An alternative string theory in twistor
space for N=4 super-Yang-Mills}, Phys. Rev. Lett. 93 (2004) 011601, 
hep-th/0402045.}
will be useful in these
computations.
\vskip 15pt

{\bf Acknowledgements:} I would like to thank  
Rajesh Gopakumar, Chris Hull, Lubos Motl, 
Nikita Nekrasov, Hirosi Ooguri, Sasha Polyakov, 
Warren Siegel, Cumrun Vafa, Brenno Carlini Vallilo,
Edward Witten, and especially Juan Maldacena for useful discussions,
CNPq grant 305814/2006-0 and 
FAPESP grant 04/11426-0
for partial financial support, and the Funda\c{c}\~ao Instituto de
F\'{\i}sica Te\'orica 
for their hospitality. I would also like to thank the organizers of
the Twistor String Theory workshop in Oxford University 
where initial stages of this research were presented at 
{\tt http://www.maths.ox.ac.uk/\char `\~ lmason/Tws/programme.html}.

\listrefs

\end